\documentclass[%
 reprint,
 amsmath,amssymb,
 aps,
prresearch,
longbibliography,
twocolumn
]{revtex4-2}

\usepackage{dcolumn}
\usepackage{bm}
\usepackage[mathlines]{lineno}
\usepackage{comment}
\usepackage[usenames]{color}
\usepackage{graphicx}
\usepackage[unicode=true,bookmarks=false]{hyperref}
\hypersetup{hidelinks}
\usepackage{braket}
\usepackage{amsthm}

\def\>{{\rangle}}
\def\<{{\langle}}
\def\bs{\boldsymbol }

\def\C{\mathcal}
\def\bb{\mathbb}

\begin{document}



\preprint{APS/123-QED}

\title{  Operator-frame geometry of non-compact quantum systems with frame-vacuum phase transitions}

\author{Satoshi Tanaka}
\email{stanaka@omu.ac.jp}
\affiliation{Department of Physics, 
Nambu Yoichiro Institute of Theoretical and Experimental Physics,
Osaka Metropolitan 
University, 3-3-138 Sugimoto Sumiyoshi-ku, Osaka-shi, 558-8585 Osaka, Japan
}
\author{Gonzalo Ordonez}
\affiliation{Department of Physics and Astronomy,
Butler University
4600 Sunset Avenue,
Indianapolis, Indiana 46208, USA} 
\author{Masatoshi Kubota}
\author{Hiroto Nakano}
\affiliation{Department of Physics, Osaka Metropolitan 
University, 3-3-138 Sugimoto Sumiyoshi-ku, Osaka-shi, 558-8585 Osaka, Japan }
\author{Bruno Mera}
\affiliation{Instituto de Telecomunica\c{c}\~oes and Departamento de Matem\'{a}tica, Instituto Superior T\'ecnico, Universidade de Lisboa, Avenida Rovisco Pais 1, 1049-001 Lisboa, Portugal}
\affiliation{Advanced Institute for Materials Research (WPI-AIMR), Tohoku University, Sendai 980-8577, Japan}
 \author{Kazuki Kanki}
\affiliation{Department of Physics, 
Nambu Yoichiro Institute of Theoretical and Experimental Physics,
Osaka Metropolitan 
University, 3-3-138 Sugimoto Sumiyoshi-ku, Osaka-shi, 558-8585 Osaka, Japan}

\date{\today}
             

\begin{abstract}
We formulate the geometric structure of non-compact bosonic quantum systems in regimes where vacuum instability renders the relevant quantum states non-normalizable, causing conventional state-space quantum geometry --- described by the Berry connection, curvature, and quantum metric --- to become ill-defined.
To overcome this breakdown, we develop a formulation of quantum geometry at the level of canonical operator frames, allowing for complexified Bogoliubov-Valatin transformations that lift the requirement that creation operators be Hermitian conjugates of annihilation operators. Canonical operator frames are defined as choices of bosonic creation and annihilation operators realizing the canonical commutation relations. A natural equivalence relation among such frames generalizes the phase ambiguity of quantum states and determines a parameter space that analytically extends the stable-regime parameter space.
The space of canonical operator frames forms a principal bundle over parameter space --- the operator-frame bundle --- equipped with a natural Ehresmann connection that defines parallel transport while preserving the canonical commutation relations. In the stable regime, this construction reduces to the Berry connection, while more generally it yields a well-defined operator-space quantum geometric tensor (QGT) that remains valid across vacuum instabilities.
Using the framework of rigged Hilbert spaces, we define a notion of quantum frame vacuum and obtain a consistent state-space QGT. Focusing on a single bosonic mode, we demonstrate analyticity of the QGT across the quantum frame-vacuum phase transition and present the corresponding phase diagram on the complexified squeezing-parameter plane. We further introduce a realistic physical setting that allows continuous paths connecting stable and unstable regimes, along which the QGT evolves smoothly across Stokes lines.

\end{abstract}

\maketitle

\section{Introduction}

The phase factor of a quantum state is one of the most characteristic features of quantum mechanics, reflecting the interpretation of the complex wavefunction in Hilbert space in terms of probability amplitudes~\cite{von1955mathematical}.
Since the phase factor represents a gauge freedom, it was long considered unobservable.
However, Berry showed that this phase becomes observable when a quantum state undergoes a cyclic evolution: the state acquires a gauge-independent geometric phase in addition to the dynamical phase, now known as the Berry phase~\cite{Berry1984b,Aharonov1987}.
At the same time, Simon provided a purely geometrical formulation of this phase in terms of fiber bundle theory~\cite{Simon1983}, showing that the Berry phase is precisely the holonomy of a connection on a Hermitian line bundle over the parameter space.
Nowadays, geometric perspectives have become ubiquitous across diverse areas of quantum mechanics, emphasizing the geometric and topological structures underlying quantum materials and  dynamics\cite{Wilczek1989,Torma2023,Ozawa2021,Mera2021}.

Recently, it has been revealed that geometric phases provide powerful tools for characterizing singular behaviors at quantum phase transitions. Carollo \textit{et al.} showed that the Berry phase exhibits singular behavior associated with the nonanalyticity of the ground state at a phase transition~\cite{Carollo2005,Carollo2020}. Subsequently, Zanardi \textit{et al.} demonstrated that the quantum geometric tensor (QGT) captures local changes of quantum geometry in the vicinity of a critical point~\cite{Cozzini2007,Zanardi2007}. The real part of the QGT defines the quantum metric tensor, which measures the Fubini-Study distance between infinitesimally close quantum states, while its imaginary part corresponds to the Berry curvature, encoding infinitesimal parallel transport with respect to the Berry connection.

These geometrical quantities are defined by means of the Hermitian inner product in Hilbert space and are applicable only as long as the relevant quantum states can be represented by vectors in a Hilbert space. However, in quantum systems with unbounded degrees of freedom, which we refer to as non-compact quantum systems, vacuum instabilities can occur at a quantum phase transition. 
In such situations, the resulting vacuum may no longer admit a normalizable representative in the Hilbert space.
 The conventional definition of the Berry connection, which relies on a Hermitian metric structure, is no longer applicable. To identify and characterize the quantum geometry of such systems, it is therefore necessary to develop a formulation of quantum geometry that does not rely on a Hermitian metric on the Hilbert space, but instead remains well defined even when the conventional state-based description breaks down.

To address this problem, we formulate quantum geometry not on the Hilbert space of states, but on a base manifold associated with canonical operator frames.
We define the quantum geometric tensor on this projective operator-frame base manifold.
Importantly, this formulation of quantum geometry does not rely on any specific Hamiltonian dynamics.
The projective operator-frame base manifold plays the role of a quantum phase-space structure that exists prior to any specific Hamiltonian dynamics.
Our aim is to elucidate the intrinsic geometric characteristics of the operator-frame bundle itself and to clarify how quantum geometry can be defined
independently of any particular dynamical model.

We organize the paper as follows.
In Sec.~\ref{sec:CM}, we first review the geometric structure of classical phase space.
Classical phase space is generically non-compact and is naturally equipped
with a symplectic structure, which provides a well-established geometric framework
independent of any specific dynamics.
For this reason, classical symplectic geometry serves as a natural reference
for developing a geometric description of non-compact quantum systems.
In Sec.~\ref{sec:Quant}, we discuss how the canonical quantization of the classical variables $(q,p)$
naturally introduces a complex structure in addition to the symplectic structure,
and we show that the local geometry of the canonical operator frame
possesses a flat K\"ahler structure.

Our central results are presented in Sec.~\ref{sec:FrameTrans},
where we establish a principal bundle structure
on the space of canonical operator frames.
An Ehresmann connection on the frame bundle identifies a horizontal tangent space,
along which the parallel transport of the canonical operators is defined; as a result, the canonical commutation relations are preserved.
A gauge-fixing condition selects a Berry connection,
which determines the associated covariant derivatives
on the total space of the operator-frame bundle.
Using the local quantum metric defined on the operator frame,
we introduce a line element associated with the infinitesimal displacement
of the parallel-transported operators.
This construction naturally defines the quantum geometric tensor
on the base manifold of the operator-frame bundle.

In Sec.~\ref{sec:FrameV}, we define the vacuum of the frame associated with each operator frame.
We show that the frame vacuum is obtained by a generally non-unitary transformation
from the bare vacuum of a reference frame $(\hat a, \hat a^\dagger)$.
The stability of the frame vacuum is controlled by a complex squeezing parameter,
and the critical boundaries that separate  stable and unstable domains
appear as Stokes lines in the complex parameter plane.
In the unstable domain, the frame vacuum no longer belongs to the Hilbert space
but must be described within the formalism of the rigged Hilbert space (RHS)~\cite{Gelfand1964,Petrosky1991a,Petrosky1997,Bohm1999,Petrosky2000,Ordonez2001,Chruscinski2003,Chruscinski2004,DeLaMadrid2005}.
We find that the QGT, defined intrinsically on the complex parameter plane,
smoothly evolves across the phase transition boundary,
whereas the frame vacuum exhibits the breakdown of the Hilbert-space description.

In Sec.\ref{sec:Physical}, we demonstrate a  physical example illustrating the continuous evolution of the QGT on the operator-frame base manifold across the Stokes lines, thereby demonstrating continuous trajectories connecting the stable and unstable domains across different Riemann sheets.
In the Appendices, we also derive the QGT for the frame eigenstates that belong to the RHS in the unstable domain.
We conclude our results in Sec.\ref{sec:Conclusion}.

\section{Symplectic Structure of the Classical Phase Space}\label{sec:CM}

In this section, we briefly review the symplectic geometry of classical mechanics, which motivates the canonical operator-frame geometry  in the quantum case to be developed in this work.
It is well known that the classical mechanics possesses a symplectic structure in the phase space, while the space of physical observables carries the algebraic structure of a Poisson algebra \cite{Kibble1978,Ashtekar1999a}.

As a simple example, we consider a classical system with a single degree of freedom,
where the phase space $M=\{(q,p)\in \mathbb{R}^2\}$ is a two-dimensional symplectic manifold $(M,\Omega_C)$ equipped with an antisymmetric, closed, and nondegenerate symplectic form:
\begin{align}\label{OmegaC1}
\Omega_C := dq \wedge dp,
\end{align}
which defines a bilinear map $\Omega_C : TM \times TM \to \mathbb{R}$.
Here the subscript $C$ denotes the classical case.

We represent the two-form $\Omega_C$ by its matrix representation
with respect to the basis $(\partial_q,\partial_p)$:
\begin{align}\label{OmegaC2}
{\bs \Omega}_C=
\begin{pmatrix}
\Omega_C(\partial_q,\partial_q) & \Omega_C(\partial_q,\partial_p)\\
\Omega_C(\partial_p,\partial_q) & \Omega_C(\partial_p,\partial_p)
\end{pmatrix}
=
\begin{pmatrix}
0 & 1\\
-1 & 0
\end{pmatrix} \;.
\end{align}
Here we refer to $(q,p)$ as a local symplectic frame, i.e., a choice of Darboux coordinates in which the symplectic form takes the canonical form \eqref{OmegaC1}.

Next we consider a dynamical trajectory on a fixed reference frame $(q_0,p_0)$,
generated by a Hamiltonian function $H$ on the phase space.
In terms of the $(q_0,p_0)$ frame, we consider a quadratic Hamiltonian:
\begin{align}\label{Hqp}
H(q_0,p_0)
&=\frac{1}{2}\left(\alpha p_0^2 + \beta q_0^2 + 2\gamma q_0 p_0 \right) \\
&=\frac 12 
\begin{pmatrix} q_0 & p_0 \end{pmatrix}
{\bs H}^{(0)}
\begin{pmatrix} q_0 \\ p_0 \end{pmatrix} \;,
\qquad (\alpha,\beta,\gamma \in \mathbb{R}) \;.
\end{align}
where ${\bs H}^{(0)}$ is the matrix representation of the quadratic Hamiltonian in the reference frame $(q_0,p_0)$, given by
\begin{align}\label{H0matrix}
{\bs H}^{(0)}=
\begin{pmatrix}
\beta & \gamma \\ \gamma & \alpha \end{pmatrix} \;.
\end{align}

The corresponding Hamiltonian vector field $X_H$ is defined by
the interior product:
\begin{align}\label{interior}
\iota_{X_H} \Omega_C = dH.
\end{align}
Since the Hamiltonian takes the quadratic form, the Hamiltonian flow is given by the canonical equation:
\begin{align}\label{Caneq}
\frac{d}{dt}
\begin{pmatrix}
q_0(t) \\
p_0(t)
\end{pmatrix}
=
{\bs X}_H^{(0)}\begin{pmatrix}q_0(t) \\p_0(t)\end{pmatrix} \;.
\end{align}
where \( {\bs X}_H^{(0)} \) denotes the matrix representation of the Hamiltonian flow in the $(q_0,p_0)$ frame:
\begin{align}\label{XH}
{\bs X}_H^{(0)}=
\begin{pmatrix}
\gamma & \alpha \\
-\beta & -\gamma
\end{pmatrix} \;.
\end{align}
Since the characteristic equation of ${\bs X}_H^{(0)}$ reads
\begin{align}
\det({\bs X}_H^{(0)}-\lambda {\bb I}_2)=\lambda^2 - \gamma^2 + \alpha\beta = 0 \;,
\end{align}
we classify the motion on the phase space $(q_0,p_0)$ as follows:
\begin{align}\label{motion}
\begin{cases}
\gamma^2 < \alpha\beta 
& \text{elliptic (closed orbits)},\\
\gamma^2 = \alpha\beta 
& \text{parabolic (separatrix)},\\
\gamma^2 > \alpha\beta 
& \text{hyperbolic (open orbits)}.
\end{cases}
\end{align}
This classification refers to the geometry of the integral curves on the reference frame.

A linear symplectic transformation from a reference frame $(q_0,p_0)$
to another frame $(q,p)$ is represented by a matrix ${\bs S}_C$
that preserves the symplectic form ${\bs \Omega}_C$:
\begin{align}\label{SympTran}
\begin{pmatrix}
q \\
p
\end{pmatrix}
=
{\bs S}_C
\begin{pmatrix}
q_0 \\
p_0
\end{pmatrix} \;,
\end{align}
where ${\bs S}_C$ is a symplectic matrix satisfying
\begin{align}\label{SclassicSymp}
{\bs S}_C^T {\bs \Omega}_C {\bs S}_C = {\bs \Omega}_C \;.
\end{align}

For a single degree of freedom, a real symplectic matrix ${\bs S}_C \in Sp(2,\mathbb{R})$
admits a Bloch--Messiah decomposition into two real orthogonal rotations
${\bs R}(\eta)$ and ${\bs R}(\theta)$, and a real hyperbolic (squeezing) transformation
${\bs B}(\beta)$ \cite{Arvind1995,Braunstein2005}:
\begin{align}\label{Sclassical}
{\bs S}_C={\bs R}(\eta) {\bs B}(\beta) {\bs R}(\theta) \;,
\end{align}
where ${\bs R}(\theta)$ and ${\bs B}(\beta)$ are given by
\begin{align}\label{SthetaC}
{\bs R}(\theta)=
\begin{pmatrix}
 \cos\theta &    -\sin\theta  \\
  \sin\theta  &     \cos\theta 
\end{pmatrix}
\;, \; \theta \in [0,2\pi)  \;,
\end{align}
and 
\begin{align}\label{SbetaC}
{\bs B}(\beta)=
\begin{pmatrix}
 \cosh\beta  &    \sinh\beta \\
  \sinh\beta &     \cosh\beta
\end{pmatrix}
\;,\; \beta \in {\mathbb R}
\end{align}
respectively.
This decomposition separates compact (rotational) and non-compact (squeezing)
degrees of freedom, which will play distinct roles in the quantum formulation.

Under the symplectic transformation \eqref{SympTran}, the Hamiltonian function itself is invariant,
but its coordinate representation changes.
In particular, for the same phase-space point,
\begin{align}
H(q_0,p_0) = H(q,p),
\qquad (q,p)^T = {\bs S}_C (q_0,p_0)^T,
\end{align}
and its matrix representation in the $(q,p)$ frame is given by
\begin{align}
H(q,p)
= \frac 12 
\begin{pmatrix} q & p \end{pmatrix}
{\bs H}^{(q,p)}
\begin{pmatrix} q \\ p \end{pmatrix} \;.
\end{align}
where
\begin{align}\label{Hqpmatrix}
{\bs H}^{(q,p)}={\bs S}_C^{-T} {\bs H}^{(0)}{\bs S}_C^{-1} \neq {\bs H}^{(0)} \;.
\end{align}
This reflects the fact that the same Hamiltonian function admits different matrix representations in different frames, while describing the same underlying dynamics.

Even so, the symplectic transformation ${\bs S}_C$ preserves the qualitative phase-space structure of the motion in different frames.
The equation of motion in the frame $(q,p)$ is obtained from \eqref{Caneq} with the symplectic transformation \eqref{SympTran}:
\begin{align}\label{transformed Ceq}
\frac{d}{dt}
\begin{pmatrix}
q(t)\\
p(t)
\end{pmatrix}
=
{\bs X}_H^{(q,p)}
\begin{pmatrix}
q(t)\\
p(t)
\end{pmatrix} \;,
\end{align}
where the Hamiltonian flow matrix is transformed from that of the reference frame by
\begin{align}
{\bs X}_H^{(q,p)} = {\bs S}_C {\bs X}_H^{(0)} {\bs S}_C^{-1} \;.
\end{align}
Since the characteristic equations coincide, i.e.,
\begin{align}
\det\!\left({\bs X}_H^{(q,p)}-\lambda {\bb I}_2\right)
=
\det\!\left({\bs X}_H^{(0)}-\lambda {\bb I}_2\right) \;,
\end{align}
the qualitative classification of the motion --- such as elliptic, parabolic, or hyperbolic behavior  --- is invariant under classical symplectic transformations.
 This confirms that a symplectic frame transformation does not alter the geometric nature of classical motion in phase space, indicating that the matrix representation of the Hamiltonian is coordinate dependent, whereas the geometric structure of the dynamics is preserved.

 This situation changes in quantum mechanics when
the classical fixed point $(q,p)=0$ is interpreted as a vacuum state.
Quantum vacuum fluctuations modify the structure of the vacuum,
and can lead to instability of the  vacuum state  once the notion of vacuum is extended beyond the Hilbert-space framework.
This phenomenon will be discussed in detail in Sec.~\ref{sec:FrameV}.

The Poisson bracket is defined via the symplectic form $\Omega_C$ evaluated on the corresponding Hamiltonian vector fields:
\begin{align}\label{Poisson}
\{f,g\} := \Omega_C(X_f,X_g) \; ,
\end{align}
where $X_f$ and $X_g$ denote the Hamiltonian vector fields associated with observable functions
$f,g \in C^\infty(M)$, respectively.
In particular, for the canonical coordinate functions $Q(q,p)=q$ and $P(q,p)=p$, one obtains
\begin{align}
\{Q,P\} = 1 \; .
\end{align}
This relation expresses a fundamental invariant of classical mechanics:  
the canonical Poisson bracket is preserved under symplectic transformations,
reflecting the underlying symplectic structure of the phase space.

\section{Geometry of operator vector space and the operator frame  structure }\label{sec:Quant}

\subsection{Local geometry of the vector space}
 
In the preceding section, we reviewed the geometrical formulation of classical
mechanics on a real symplectic manifold $(M,\Omega_C)$ for a single degree of
freedom, with $M=\mathbb{R}^2$.  
Upon quantization, the classical canonical variables $(q,p)$ are promoted to the self-adjoint operators $(\hat q,\hat p)$ satisfying the canonical commutation relation $[\hat q, \hat p]=i\hbar$.

To introduce a complex operator basis on the operator space,
we perform a linear canonical transformation from the self-adjoint
canonical pair $(\hat q, \hat p)$ to the complex canonical pair
$(\hat a, \hat a^\dagger)$:
\begin{align}\label{unitaryTransa}
\hat{\bs\phi}_0:= \begin{pmatrix}
\hat a \\
\hat a^\dagger
\end{pmatrix}
=
{\bs U}
\begin{pmatrix}
\hat q \\
\hat p
\end{pmatrix},
\qquad
{\bs U}=\frac{1}{\sqrt 2}
\begin{pmatrix}
1 & i \\
1 & -i
\end{pmatrix},
\end{align}
which yields the normalized commutation relation $[\hat a,\hat a^\dagger]=1$.  

Rather than working in the Hilbert space of states, we formulate the geometry in the operator vector space spanned by the complex canonical operators.
We define the \emph{operator vector space}
\begin{align}\label{V}
V = \mathrm{span}_{\bb{C}}\{\hat a,\hat a^\dagger\}\;.
\end{align}
Note that the pair $(\hat a,\hat a^\dagger)$ in $\hat{\bs\phi}_0$ serves as an ordered basis (reference frame) of the operator vector space $V$. 
Different canonical operator pairs correspond to different ordered bases
of the same operator vector space $V$.

While in classical mechanics the symplectic form acts on tangent vectors, in the operator space $V$ we instead define the symplectic structure directly on operators themselves.
We define an antisymmetric, non-degenerate  bilinear form on $V$ by  
\begin{align}\label{OmegaV}
&\Omega: V\times V \to {\bb C} \notag\\
&\Omega(\hat\psi,\hat\psi') := [\hat\psi,\hat\psi'] \qquad (\hat\psi,\hat\psi'\in V) \;.
\end{align}

The transformation from the real canonical variables $(q,p)$ to the complex
operator pair $(\hat a,\hat a^\dagger)$ motivates the introduction of
a complex structure on $V$.
The complex structure $J_0$ is defined as a linear endomorphism on $V$ satisfying
\begin{align}\label{Jmap}
J_0: V \to V, \qquad J_0^2=-\mathrm{id}_V\;,
\end{align}
and the compatibility with the symplectic structure:
\begin{align}
\Omega(J_0(\hat\psi), J_0(\hat\psi'))
=
\Omega(\hat\psi,\hat\psi'),
\qquad
(\hat\psi,\hat\psi'\in V)\;.
\end{align}
In terms of the reference frame $\hat{\bs\phi}_0$, we specifically define $J_0$ on $V$ as 
\begin{align}\label{JonV}
J_0(\hat a)=\hat a^\dagger,\qquad
J_0(\hat a^\dagger)=-\hat a \;.
\end{align}
Note that the specific action of $J$ depends on the operator frame as shown in Appendix~\ref{AppSec:Complex}.

Together with the symplectic form $\Omega$ and the complex structure $J_0$, we  define a symmetric bilinear form on $V$  by
\begin{align}\label{GV}
& G_0: V\times V\to {\bb C} \notag\\
&G_0(\hat\psi,\hat\psi')=\Omega(\hat\psi, J_0(\hat\psi')),\qquad(\hat\psi,\hat\psi'\in V)\;.
\end{align}
We note that the metric structure generally depends on the choice of local operator frame through the corresponding complex structure.
In \eqref{GV},  $(G,J,\Omega)$ constitutes a K\"ahler-type structure on the operator space $V$,
providing a metric structure that will play a central role in defining the quantum geometric tensor,
as shown in Sec.~\ref{sec:FrameTrans}.

\subsection{Frame  transformation}

We now introduce the symplectic operator frame space generated by the complex symplectic transformations of the canonical operators, together with the fiber structure induced by the gauge symmetry.
Associated with the operator vector space $V$, we consider the symplectic frame space ${\rm Fr}_\Omega(V)$ given by the collection of the  ordered bases in $V$ under the constraint of the canonical commutation relation: 
\begin{align}\label{FrameV}
{\rm Fr}_\Omega(V)=
\left\{(\hat\varphi,\hat\varphi^\star)\in V\times V \ \middle| \  [\hat\varphi,\hat\varphi^\star]=1 \right\} \;.
\end{align}

In terms of the reference frame $\hat{\bs\phi}_0$ , the two-form $\Omega$ has a matrix representation: 
\begin{align}\label{OmegaV2}
{\bs \Omega}(\hat{\bs\phi}_0):=
\begin{pmatrix}
\Omega(\hat a,\hat a) & \Omega(\hat a,\hat a^\dagger) \\
\Omega(\hat a^\dagger,\hat a) & \Omega(\hat a^\dagger,\hat a^\dagger)
\end{pmatrix}=
\begin{pmatrix}
0 & 1 \\ -1 & 0
\end{pmatrix},
\end{align}
whose matrix form coincides with that of the classical symplectic form
\eqref{OmegaC2}.
The action of $J_0$ on the reference frame is then defined as  a map 
\begin{align}\label{JonFr0}
&\bs J_0 : {\rm Fr}_\Omega(V) \to {\rm Fr}_\Omega(V)  \;,\notag\\
&{\bs J}_0(\hat{\bs \phi}_0) :=
\begin{pmatrix}
J_0(\hat a) \\
J_0(\hat a^\dagger)
\end{pmatrix}
=\begin{pmatrix}
\hat a^\dagger \\
-\hat a
\end{pmatrix} 
=
\begin{pmatrix}
0 & 1\\
-1 & 0
\end{pmatrix}
\begin{pmatrix}
\hat a \\
\hat a^\dagger
\end{pmatrix} \;,
\end{align}
where the compatibility condition of $J$ with the symplectic form ensures that
the transformed frame also satisfies the canonical commutation relation.
Therefore, in the reference frame $\hat{\bs\phi}_0$, the complex structure
admits the constant matrix representation
\begin{align}\label{Jmat}
\bs J_0=
\begin{pmatrix}
0 & 1\\
-1 & 0
\end{pmatrix}  \;.
\end{align}
In terms of  the reference frame $\hat{\bs\phi}_0$, $G_0$ has the matrix representation:
\begin{align}
{\bs G}_0(\hat{\bs\phi}_0):=\begin{pmatrix}
G_0(\hat a,\hat a) & G(\hat a,\hat a^\dagger) \\
G_0(\hat a^\dagger,\hat a) &G(\hat a^\dagger,\hat a^\dagger)
\end{pmatrix}
=\begin{pmatrix} 1 & 0 \\ 0 & 1 \end{pmatrix} \;,
\end{align}
showing that the metric is flat in the reference frame.

Analogous to the classical symplectic transformation \eqref{SympTran},
the  reference operator frame $\hat{\bs\phi}_0=(\hat a,\hat a^\dagger)^T$
is  transformed into another ordered operator frame
$\hat{\bs\phi}_\xi=(\hat\varphi_\xi,\hat\varphi_\xi^\star)^T $ by a complex symplectic transformation.

We define a symplectic frame transformation as a map
\begin{align}\label{SmaponFr}
{\bs S}:{\rm Fr}_\Omega(V)\to{\rm Fr}_\Omega(V).
\end{align}
The transformation is represented by a matrix
${\bs S}_\xi$ acting on the reference frame as
\begin{align}\label{Sactonphi0}
\hat{\bs\phi}_\xi=
\begin{pmatrix}
\hat\varphi_\xi\\
\hat\varphi_\xi^\star
\end{pmatrix}
={\bs S}_\xi (\hat{\bs\phi}_0)
=\bs S_\xi
\begin{pmatrix}
\hat a \\
\hat a^\dagger
\end{pmatrix} \;,
\end{align}
where $\xi$ collectively denotes the parameters characterizing the symplectic transformation. 
As discussed below, one combination of these parameters corresponds to the gauge direction associated with ${\mathbb C}^\times$.
Since the symplectic transformation preserves the canonical commutation relation, we verify that the symplectic form remains invariant on the different frames:
\begin{align}\label{phiCCR}
&{\bs\Omega}(\hat{\bs\phi}_\xi)=
\begin{pmatrix}
\Omega(\hat\varphi_\xi,\hat\varphi_\xi) & \Omega(\hat\varphi_\xi,\hat\varphi^\star_\xi) \\
\Omega(\hat\varphi^\star_\xi,\hat\varphi_\xi) & \Omega(\hat\varphi^\star_\xi,\hat\varphi^\star_\xi) 
\end{pmatrix} \notag\\
&=\begin{pmatrix}
[\hat\varphi_\xi,\hat\varphi_\xi] & [\hat\varphi_\xi,\hat\varphi^\star_\xi]\\
[\hat\varphi^\star_\xi,\hat\varphi_\xi] &[\hat\varphi^\star_\xi,\hat\varphi^\star_\xi]
\end{pmatrix}
=
\begin{pmatrix} 0 & 1 \\ -1 & 0 
\end{pmatrix}
=
\bs\Omega(\hat{\bs\phi}_0) \;,
\end{align}
where $\bs\Omega(\hat{\bs\phi}_0)$ is given in \eqref{OmegaV2}.
This leads to the matrix relation
\begin{align}\label{OmegaPreserve}
{\bs S}_\xi^T \bs\Omega {\bs S}_\xi = \bs\Omega.
\end{align}
\begin{proof}
For simplicity, we drop the suffix $\xi$ in the following proof.
 Since $\bs S$ is represented by a $2\times 2$-matrix seen from \eqref{Sactonphi0},  we represent the linear transformation of the operators as 
\begin{align}\label{SxionV}
\begin{pmatrix}
\hat\varphi \\
\hat\varphi^\star
\end{pmatrix}
=\begin{pmatrix}
S_{1,1} & S_{1,2} \\
S_{2,1} & S_{2,2}
\end{pmatrix}
\begin{pmatrix}
\hat a \\
\hat a^\dagger
\end{pmatrix} \;.
\end{align}
We represent this transformation by $\hat\psi_i=S_i^j \hat c_j$  using the Einstein summation convention, where  we denote $\hat\psi_1=\hat\varphi$, $\hat\psi_2=\hat\varphi^\star$, $\hat c_1=\hat a$, and $\hat c_2=\hat a^\dagger$ with  the indices $i (, j)=1$ and $2$.
Then, Eq.~\eqref{OmegaPreserve} follows as $ \bs \Omega_{ij}=\Omega(\hat\psi_i,\hat\psi_j)=\Omega( S_i^l \hat c_l  ,  S_j^m \hat c_m)=S_i^l  S_j^m   \Omega(\hat c_l,\hat c_m) 
=S_i^l    \bs\Omega_{l m} S_j^m  
= ( \bs S^T  \bs \Omega \bs S)_{ij} $ \;.
\end{proof}

In practice, the complex symplectic transformation ${\bs S}_\xi$ is represented by a matrix product \cite{Braunstein2005}:
\begin{align}
{\bs S}_\xi = {\bs S}_\eta {\bs S}_\beta {\bs S}_\theta,
\end{align}
where these matrices are given by
\begin{align}
&\bs S_{\eta}
=\begin{pmatrix}
e^{i\eta} & 0 \\
0 & e^{-i\eta}
\end{pmatrix}
\;,\;
\bs S_{\beta}
=\begin{pmatrix}
\cosh\beta & \sinh\beta \\
\sinh\beta & \cosh\beta
\end{pmatrix}
\;,\; \notag\\
&\bs S_{\theta}
=\begin{pmatrix}
e^{i\theta} & 0 \\
0 & e^{-i\theta}
\end{pmatrix} \;,\;  ( \eta,\theta \in {\bb C},\beta\in {\bb C}^\times  ) \;.
\end{align}
In Appendix~\ref{AppSec:Complex}, we show the transformation property of the complex structure under the symplectic transformation.

Since there is a gauge symmetry associated with an element of ${\rm Fr}_\Omega(V)$:
\begin{align}
(\hat\varphi,\hat\varphi^\star)\sim (\lambda \hat\varphi, \lambda^{-1}\hat\varphi^\star)\;, \; \lambda \in {\bb C}^\times  \;,
\end{align}
we  consider the projective operator-frame manifold as the base manifold $B$ given by
\begin{align}
B={\rm Fr}_\Omega(V)/{\bb C}^\times =\{ [\hat{\bs\phi}]  |  \hat{\bs\phi} \in {\rm Fr}_\Omega(V) \} \;.
\end{align}
Then we obtain the principal ${\mathbb C}^\times$-bundle
\begin{align}
\mathbb{C}^\times \hookrightarrow {\rm Fr}_\Omega(V)
\to {\rm Fr}_\Omega(V)/\mathbb{C}^\times.
\end{align}
This construction defines a principal bundle structure.

\section{Berry connection and quantum geometric tensor}\label{sec:FrameTrans}

\subsection{Ehresmann connection on the principal bundle}

In the present operator-frame principal bundle, the conventional Hilbert-space definition of the
Berry connection, which relies on a Hermitian inner product, is not applicable.
For this reason, we introduce an Ehresmann connection, which is applicable
independently of the compactness of the structure group $\mathbb{C}^\times$ and does not rely
on any inner product; see Appendix~\ref{AppSec:Ehresmann}
\cite{kobayashi1996foundations,morita2001geometry,NakaharaBook}.

We identify the symplectic operator-frame space 
\begin{align}
P := {\rm Fr}_\Omega(V)
\end{align}
as the total space of a principal $\mathbb{C}^\times$-bundle over the base manifold $B$,
with projection $\pi:P\to B$.

An Ehresmann connection on $P$ is defined as a decomposition of the tangent space
at each point $u\in P$ into horizontal and vertical subspaces:
\begin{align}
T_u P = H_u P \oplus V_u P.
\end{align}
The vertical subspace $V_u P$ is given by the kernel of the differential of the projection,
\begin{align}
V_u P = \ker (d\pi_u),
\end{align}
and is generated by the fundamental vector fields associated with the Lie algebra
$\mathfrak{g} = {\rm Lie}(\mathbb{C}^\times)$.
In the present setting, the vertical directions correspond to infinitesimal
complex gauge transformations of the operator frame,
$(\hat\varphi,\hat\varphi^\star)\mapsto(\lambda\hat\varphi,\lambda^{-1}\hat\varphi^\star)$,
with $\lambda\in\mathbb{C}^\times$.
The horizontal subspace $H_u P$ is defined by the choice of connection and specifies
how operator frames are transported along the base manifold.

Equivalently, the Ehresmann connection can be described by a connection one-form
\begin{align}
\omega \in \Omega^1(P,\mathfrak{g}),
\end{align}
which projects tangent vectors onto the vertical subspace associated with the Lie algebra action.

A lifted curve $u(t)\in P$ is said to be horizontal if its tangent vector satisfies
\begin{align}
\omega(\dot u(t)) = 0,
\end{align}
which defines the horizontal lift of curves on the base manifold and thereby induces the parallel transport of operator frames.
This construction is independent of any metric structure and is therefore applicable
to the present setting with the non-compact structure group $\mathbb{C}^\times$.
In the following, this geometric notion of horizontal transport will be realized
in terms of a covariant derivative acting on operator frames.

\subsection{The covariant derivatives on the associated vector bundle}

In this subsection, we derive the covariant derivative on the associated vector bundle
\begin{align}
E = P \times_\rho V,
\end{align}
constructed from the principal ${\bb C}^\times$-bundle $P$ and the vector space $V$.
The action of the structure group $\mathbb C^\times$ on $V$ is specified by
\begin{align}
\rho:\mathbb C^\times\to GL(V).
\end{align}
We represent an element of the structure group by $\lambda\in\mathbb C^\times$
and adopt
\begin{align}
\rho(\lambda)
=
{\rm diag}(\lambda,\lambda^{-1}),
\end{align}
which reduces to
$\rho(e^{i\eta})={\rm diag}(e^{i\eta},e^{-i\eta})$
in the stable $U(1)$ regime.

Choosing a local section $s:B\to P$, corresponding to the gauge choice $\eta=0$, a curve $\boldsymbol{x}(t)$ on the base manifold
is lifted to a horizontal curve $\tilde{\gamma}(t) \in P$.
This induces operator-valued local sections of the associated bundle $E$.
These sections define a local operator frame
\begin{align}
\hat{\bs\phi}(\bs{x})
:= \hat{\bs\phi}_e
=
\begin{pmatrix}
\hat\varphi_e(\boldsymbol{x})\\
\hat\varphi_e^\star(\boldsymbol{x})
\end{pmatrix} \in {\rm Fr}_\Omega (E_{\bs x}),
\end{align}
where ${\rm Fr}_\Omega (E_{\bs x})$ denotes the space of local operator frames  over ${\bs x}=(\theta,\beta) \in B$. 
The operator-frame bundle is defined by
\begin{align}
{\rm Fr}_\Omega(E):=\bigsqcup_{\bs x \in B}{\rm Fr}_\Omega(E_{\bs x}) \;.
\end{align}
In the following, all geometric quantities are evaluated on this local operator frame,
and the corresponding K\"ahler-type structures $J_e$ and $G_e$
are understood as the pullback of the operator-space geometry to this section.

We now introduce the covariant derivative $\bs D_\mu$ acting on the operator frame,
constructed from the component covariant derivatives
$D_\mu$ and $D_\mu^\star$ acting on
$\hat\varphi_\xi$ and $\hat\varphi_\xi^\star$,
respectively.
Using the gauge potential $A_\mu(\boldsymbol{x})$,
the covariant derivative on the frame is defined as
\begin{align}\label{CovariantD}
{\bs D}_\mu({\bs x})
:=
\begin{pmatrix}
D_\mu ({\bs x}) & 0\\
0 & D^\star_\mu ({\bs x})
\end{pmatrix}
=
\begin{pmatrix}
\partial_\mu + i A_\mu({\bs x}) & 0 \\
0 & \partial_\mu - i A_\mu({\bs x})
\end{pmatrix}.
\end{align}

Using the covariant derivatives, we define the horizontally transported operator frame from $\bs x$ to $\bs x+d\bs x$ by
\begin{align}
\hat{\bs\phi}_e(\bs x+d\bs x):=
\exp[dx^\mu \bs D_\mu]\hat{\bs\phi}_e(\bs x) \;.
\end{align}
Here $\hat{\bs\phi}_e(\bs x+d\bs x)$ denotes the horizontal representative obtained by parallel transport from $\bs x$, rather than an independently chosen value of a local section at the neighboring point.
The components
$\hat{\bs\phi}_e(\bs x+d\bs x)=(\hat\varphi_e(\bs x+d\bs x),\hat\varphi_e^\star(\bs x+d\bs x))^T$
are given by the covariant Taylor expansion in $dx^\mu$ \cite{Avramidi1999}:
\begin{subequations}\label{CovTaylor}
\begin{align}
\hat\varphi_e(\bs x + d\bs x) :=& e^{dx^\mu D_\mu}\hat\varphi_e(\bs x)\notag\\
=&\hat\varphi_e(\bs x) + D_\mu \hat\varphi_e(\bs x)\, dx^\mu \notag\\
&+ \frac{1}{2} D_{(\mu} D_{\nu)} \hat\varphi_e(\bs x)\, dx^\mu dx^\nu + O(d\bs x^3) \;,\\
\hat\varphi^\star_e(\bs x + d\bs x) :=&e^{dx^\mu D^\star_\mu}\hat\varphi^\star_e(\bs x)\notag\\
=& \hat\varphi_e^\star(\bs x) + D_\mu^\star \hat\varphi_e^\star(\bs x)\, dx^\mu\notag\\
& + \frac{1}{2} D^\star_{(\mu} D^\star_{\nu)} \hat\varphi_e^\star(\bs x)\, dx^\mu dx^\nu + O(d\bs x^3)\;,
\end{align}
\end{subequations}
where   $D_{(\mu}D_{\nu)}$ denotes symmetrization over the indices.
We have proved in Appendix~\ref{AppSec:CCR} that this parallel transport preserves the canonical commutation relation exactly:
\begin{align}\label{CCRparallel}
[\hat\varphi_e(\bs x+d\bs x),\hat\varphi_e^\star(\bs x+d\bs x)] = 1\;,
\end{align}
ensuring that the operator algebra is invariant under the parallel transport.

\subsection{Quantum geometric tensor on the operator-frame space}

Now we consider the infinitesimal displacement  of the operator-space frame under the covariant derivatives such that
\begin{align}
\delta{\hat{\bs \phi}}_e(\bs x)&:=\hat{\bs \phi}_e(\bs x+d\bs x)-\hat{\bs \phi}_e(\bs x) \notag\\
&=(\delta\hat\varphi_e(\bs x),\delta\hat\varphi_e^\star(\bs x))^T \in T_{\hat{\bs\phi}_e(\bs x)}{\rm Fr}_\Omega(E)\;,
\end{align}
where the infinitesimal parallel displacement $\delta\hat\varphi_e(\bs x) $ along $\bs x$ on $B$ is given by
\begin{align}\label{deltavarphi}
\delta\hat\varphi_e(\bs x):=
\hat\varphi_e(\bs x+d\bs x)-\hat\varphi_e(\bs x).
\end{align}

In our work, as a measure of the distance between the frames, we consider the quadratic form associated with $\delta\hat{\bs\phi}_e$:
\begin{align}\label{ds2}
ds^2&:=\frac 12 \left(G_e(\delta\hat\varphi_e(\bs x), \delta\hat\varphi_e(\bs x))+G_e(\delta\hat\varphi^\star_e(\bs x), \delta\hat\varphi^\star_e(\bs x))\right) \notag\\
&=G_e(\delta\hat\varphi_e(\bs x), \delta\hat\varphi_e(\bs x)) \;.
\end{align}
By the covariant Taylor expansion \eqref{CovTaylor}, $ds^2$ is written up to second order in $d\bs x$,
\begin{align}\label{ds2GV}
ds^2
=
G_e(D_\nu \hat\varphi_e(\bs x),D_\mu \hat\varphi_e(\bs x))
\, dx^\mu dx^\nu
+ O(d\bs x^3).
\end{align}
With the same spirit of the definition of the quantum geometric tensor from the Fubini-Study distance\cite{Provost1980}, we define the quantum geometric tensor (QGT) in the operator space as
\begin{align}\label{QGTdef}
Q_{\mu\nu}(\bs x):=-G_e(D_\nu \hat\varphi_e(\bs x),D_\mu \hat\varphi_e(\bs x)) \;,
\end{align}
where the minus sign is taken so that the quantum metric becomes positive in the stable case as shown below.
Using the K\"ahler structure on the local operator frame defined in (\ref{GV}), \eqref{QGTdef} is rewritten by 
\begin{align}\label{QGTB}
Q_{\mu\nu}({\bs x})
:=[D_\mu^{\star}\hat\varphi_e^{\star}({\bs x}), D_\nu\hat\varphi_e({\bs x})] \;,
\end{align}
as proved in Appendix\ref{AppSec:Intertwining}.
This ordering is chosen so that the symmetric part of $Q_{\mu\nu}$
reduces to a positive-definite quantum metric in the stable regime.

In order to specify the Berry connection,  we further impose the gauge fixing condition given by
\begin{align}\label{gaugeFix1}
[D_\mu\hat\varphi_e, \hat\varphi_e^\star]=0 \;,
\end{align}
which automatically ensures  
\begin{align}\label{gaugeFix2}
[\hat\varphi_e, D_\mu^\star\hat\varphi_e^\star]=0 \;,
\end{align}
because of the identity
\begin{align}
[D_\mu\hat\varphi_e, \hat\varphi_e^\star]+[\hat\varphi_e, D_\mu^\star\hat\varphi_e^\star]=0 \;.
\end{align}
Equations \eqref{gaugeFix1} and \eqref{gaugeFix2}
fix the gauge and determine the Berry connection in the operator space as
\begin{align}\label{BerryA}
A_\mu({\bs x})=i[\partial_\mu\hat \varphi_e ^\star({\bs x}), \hat \varphi_e({\bs x})  ]\;,
\end{align}
which explicitly gives
\begin{align}
A({\bs x})=(A_\theta,A_\beta)=\left(-\cosh(2\beta),0\right).
\end{align}
This result follows directly from \eqref{gaugeFix1} and the definition of $D_\mu$.
Substituting \eqref{CovariantD} and \eqref{BerryA} into \eqref{QGTB}, we obtain a specific form of the QGT as
\begin{align}\label{QGTBform}
 Q_{\mu\nu}({\bs x})&=[\partial_\mu \hat\varphi_e  ^\star({\bs x}),\partial_\nu \hat\varphi_e({\bs x})  ] \notag\\
&-[\partial_\mu\hat \varphi  ^\star_e({\bs x}), \hat\varphi_e({\bs x})  ][\partial_\nu \hat\varphi_e({\bs x})  , \hat\varphi  ^\star_e({\bs x})] \;, ({\bs x} \in B ) \;.
\end{align}
This tensor  corresponds to the conventional QGT defined in Hilbert space~\cite{Provost1980,Zhang2024}.
In contrast to the conventional derivation in Hilbert space,
the present formulation shows that the QGT arises solely
from the intrinsic geometry of the operator frame bundle,
without any reference to quantum states or wavefunctions.

For a general QGT, the tensor decomposes into a symmetric part, representing the quantum metric tensor, and an antisymmetric part, representing the Berry curvature tensor: $Q_{\mu\nu}=g_{\mu\nu}+(i/2) F_{\mu\nu}$, where $g_{\mu\nu}$ and $F_{\mu\nu}$ are defined respectively as the symmetric and antisymmetric parts of $Q_{\mu\nu}$.
In the present formulation, using the base-manifold coordinates $(\theta, \beta)$ on $B$, the two tensors take the following explicit forms:
\begin{align}\label{gmunu}
g_{\mu\nu}:={1\over 2}\left(Q_{\mu\nu}+Q_{\nu\mu}\right)
=
\begin{pmatrix}
\sinh^2 2\beta &0 \\
0 & 1
\end{pmatrix} 
\end{align}
and 
\begin{align}\label{Fmunu}
F_{\mu\nu}:=-i (Q_{\mu\nu}-Q_{\nu\mu})=
\begin{pmatrix}
0 &  2\sinh 2\beta \\
-2\sinh 2\beta  & 0
\end{pmatrix} \;,
\end{align}
where the sign convention follows our previous definition.

The Berry curvature tensor $F_{\mu\nu}$ defines the symplectic two-form $\Omega_B: = (1/2)F_{\mu\nu}dx^\mu\wedge dx^\nu$, while the quantum metric yields the bilinear form $G_B:= g_{\mu\nu}dx^\mu\otimes dx^\nu$ on the base manifold $B$.
From these, we  construct the $(1,1)$ tensor $w^\nu_\mu := \Omega_{\mu\sigma} g^{\sigma\nu}$, which satisfies $w^\nu_\sigma w^\sigma_\mu =-\delta^\nu_\mu$.
Hence $W_B := w^\nu_\mu dx^\mu\otimes\partial_\nu$ defines an almost complex structure on the base manifold and fulfills the K{\"a}hler compatibility condition 
\begin{align}\label{Bcompati}
G_B(X,Y)=\Omega_B(X,W_B Y)  , \quad ^\forall X,Y \in TB \;.
\end{align}
Here  we note that the base-manifold coordinates $(\theta,\beta)$ are not restricted to be real,
but are regarded as complex variables.
Accordingly, the quantum metric $g_{\mu\nu}$ and the Berry curvature $F_{\mu\nu}$
are generally complex-valued tensors, while the algebraic K\"ahler compatibility
conditions remain valid.

Importantly, these tensors and the associated geometric structures are defined
on the projective operator-space manifold, which serves as the base manifold
of the principal frame bundle.
Each local operator frame $\hat{\bs\phi}_\xi$ induces a locally defined K\"ahler-type structure
$(G_\xi,J_\xi,\Omega)$ on the operator space.
While the symplectic form $\Omega$ is common to all frames,
the complex structure $J_\xi$ and the associated metric $G_\xi$
are induced in each frame through the corresponding symplectic transformation.
For any fixed frame, the triple $(G_\xi,J_\xi,\Omega)$ is locally flat
and can be chosen in the same canonical form.
The nontrivial geometry therefore emerges on the base manifold through the connection of the principal frame bundle.
Moreover, since the QGT in Eqs.~\eqref{gmunu} and \eqref{Fmunu}
is analytic in $\beta$, it admits analytic continuation across
the frame-vacuum phase-transition boundaries (Stokes lines), which will be discussed in the next section.
Thus, the QGT provides a natural geometric probe of transitions between stable and unstable domains.

The conventional definition of the quantum geometric tensor (QGT) on the
projective Hilbert space was originally introduced through the Fubini-Study
distance and subsequently made gauge invariant \cite{Provost1980}.
It was later shown that this construction can be reformulated geometrically
in terms of Ehresmann connections on principal bundles with compact structure
groups \cite{Zanardi2007,Kolodrubetz2013}.
Such constructions, however, rely crucially on the Hermitian inner product structure of the underlying Hilbert space.
These approaches cannot be directly applied to the operator-space manifold
considered here, since no natural Hermitian inner product structure is available on the operator space,  and the relevant structure group is the non-compact group
$\mathbb{C}^\times$.

In contrast, the present formulation derives the QGT directly from an
Ehresmann connection on the operator-frame principal bundle, without invoking
any inner product in the Hilbert space.
Within this framework, the geometric structures are not postulated \emph{a priori}, but are
systematically constructed from the connection formalism itself, providing a
logically consistent and manifestly covariant foundation for quantum geometry.

It should be emphasized that, in many applications in condensed-matter physics,
quantum geometric quantities such as the Berry connection and the QGT are
introduced after specifying a Hamiltonian, and are analyzed in terms of its
eigenstates or dynamical evolution.
In such approaches, the geometric structures naturally appear as properties
associated with a given Hamiltonian.
In contrast, the present formulation suggests that the relevant geometric
structure is already encoded in the operator-space and frame-bundle structure,
independently of any particular choice of Hamiltonian.
In this sense, quantum geometry may be viewed as a background structure
upon which different dynamical models can be formulated,
rather than as a quantity derived solely from a specific Hamiltonian.


\section{Frame vacuum instability}\label{sec:FrameV}

In this section we define the quantum vacuum associated with each local frame,
which we call the \textit{frame vacuum}.
While the operator-space geometry constructed in the previous section
remains well defined for complex values of $\beta$,
the corresponding frame vacuum ceases to be normalizable
at certain critical boundaries in the $\beta$-plane,
known as Stokes lines~\cite{BenderBook}.

The \textit{right-frame vacuum} $|\varphi_{\xi,0}\>$ for a local frame $(\hat\varphi_{\xi},\hat\varphi_{\xi}^{\star})$ is defined as the state annihilated by the action of  $\hat\varphi_\xi$:
\begin{align}\label{vacuumdef}
\hat\varphi_\xi\,|\varphi_{\xi,0}\> = 0.
\end{align}
In order to obtain an explicit representation of  $|\varphi_{\xi,0}\>$, 
it is convenient to express the complex symplectic 
transformation in terms of exponential one-parameter operators 
generated by the $\mathfrak{su}(1,1)$ algebra:
\begin{align}
\hat V_j(\zeta_j) := e^{ i \zeta_j \hat K_j},
\qquad 
\zeta_j \in \mathbb C, 
\qquad j=1,2,3 ,
\end{align}
where $\hat K_j$ $(j=1,2,3)$ are the generators satisfying the ${\mathfrak {su}}(1,1)$ Lie-algebra:
\begin{align}\label{Ks-def}
\hat K_3 &= \frac 12 \left(\hat a \hat a^{\dagger} 
+ \hat a^{\dagger} \hat a \right), 
\qquad
\hat K_1 = \frac 12 \left( \hat a^{2} + \hat a^{\dagger 2} \right),
\notag \\
\hat K_2 &= \frac{i}{2}\!\left( \hat a^{2} - \hat a^{\dagger 2} \right)\;.
\end{align}

With these operators, the action of the symplectic transformation on 
the reference frame can be written as
\begin{align}\label{operatorRep}
\begin{pmatrix}
\hat\varphi_{\xi} \\
\hat\varphi_{\xi}^{\star}
\end{pmatrix}
=
\hat V_3^{-1}(\theta)\,
\hat V_2^{-1}(\beta)\,
\hat V_3^{-1}(\eta)
\begin{pmatrix}
\hat a \\
\hat a^{\dagger}
\end{pmatrix}
\hat V_3(\eta)\,
\hat V_2(\beta)\,
\hat V_3(\theta),
\end{align}
where $\theta,\beta,\eta$ denote the transformation 
parameters defined in \S~\ref{sec:FrameTrans}.  
Using (\ref{operatorRep}) and  $\hat  a |0\> = 0$,  
we obtain 
\begin{align}\label{frame-Rvac}
|\varphi_{\xi,0}\rangle
=
e^{-i\eta/2}\,
\hat V_3^{-1}(\theta)\,
\hat V_2^{-1}(\beta)\,
|0\rangle ,
\end{align}
where $|0\>$ denotes the reference vacuum.
The factor $e^{-i\eta/2}$ arises from the gauge generator  $\hat V_3^{-1}(\eta)$ but contributes only an overall phase.

To explore the global structure of the operator-frame space, we parametrize the squeezing
parameter as a complex variable $\beta = u + i v$ with $(u,v)\in\mathbb{R}^2$.
In general, $\theta$ may also be complex, in which case the base manifold 
becomes a complex two-dimensional (real four-dimensional) manifold 
that can be identified with the quotient manifold 
$Sp(2,\mathbb{C})/\mathbb{C}^\times \simeq SL(2,\mathbb{C})/\mathbb{C}^\times$.
In this work, however, we restrict $\theta$ to be real, since it does not induce mixing between the annihilation and creation operators and therefore is not directly related to the frame-vacuum instability.
As a result, the physically relevant base manifold considered here is a
three-dimensional real submanifold parametrized by
$(\theta,\mathrm{Re}\,\beta,\mathrm{Im}\,\beta)$.

\begin{figure}[t]
\begin{center}
\includegraphics[width=7cm]{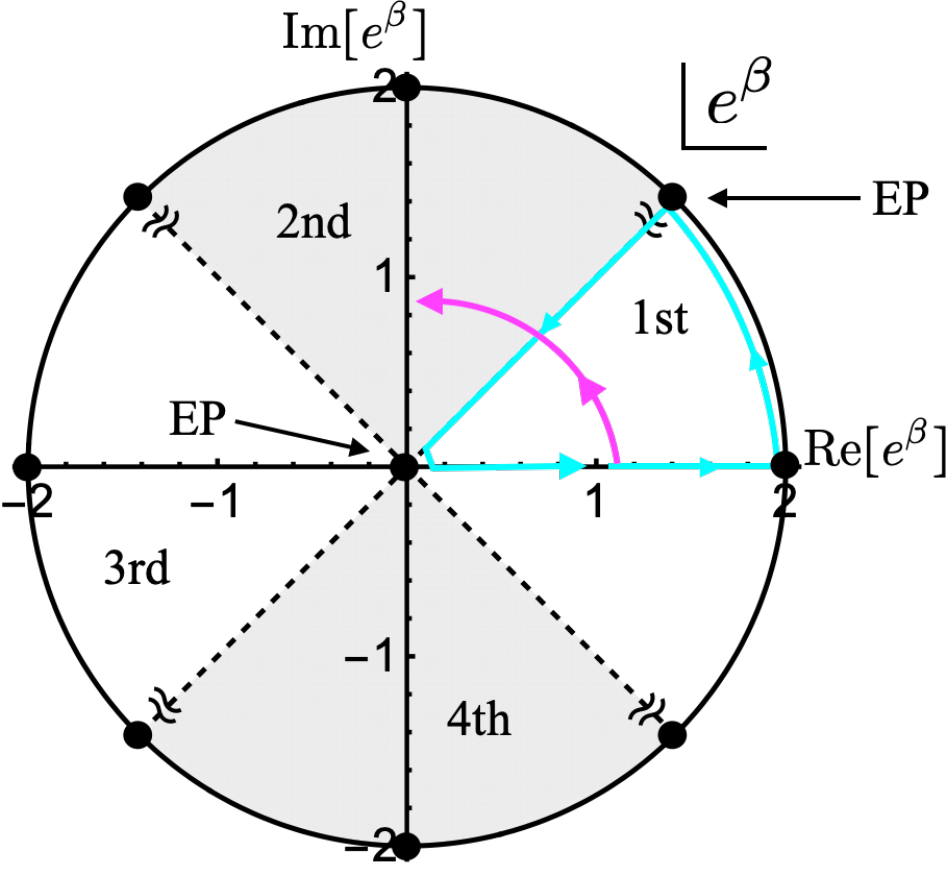}
\caption{Stable-unstable domain phase diagram in the complex $e^{\beta}$ plane.
The blank regions represent stable domains in which the frame vacuum is normalizable, while the shaded regions correspond to unstable domains where the Hilbert-space norm is no longer real positive.
The norm diverges on the Stokes lines.
The dotted lines indicate the Stokes lines separating different analytic sectors.
The exceptional points are indicated by the black dots.
The Stokes lines divide the complex plane into four Riemann sheets, labeled in the figure.
Representative trajectories connecting stable and unstable domains are also shown (blue and magenta curves); the physical realization constructing the trajectories is discussed in Sec.~\ref{sec:Physical}.
}
\label{fig:Domain}
\end{center}
\end{figure}

Using \eqref{frame-Rvac}, we evaluate the norm of the frame vacuum
in the stable region as
\begin{align}\label{varphiNorm}
\< \varphi_{\xi,0}|\varphi_{\xi,0}\>
&= e^{i(\bar\eta-\eta)/2}
   \< 0|(\hat V_2^{-1}(\beta))^\dagger \hat V_2^{-1}(\beta)|0\>
   \notag\\
&= e^{\mathrm{Im}\,\eta}
   \< 0|e^{i(\bar\beta-\beta)\hat K_2}|0\>
   \notag\\
&= e^{\mathrm{Im}\,\eta}
   \< 0|e^{2v\hat K_2}|0\>
 = \frac{e^{\mathrm{Im}\,\eta}}{\sqrt{\cos(2v)}} \;,
\end{align}
where the last equality follows from the standard vacuum-squeezed-vacuum
overlap formula obtained via the $SU(1,1)$ disentangling identity
\cite{barnett2002methods}.
Although the gauge function $\eta$ can become complex in the stable region, defined as the region surrounded by the Stokes lines shown in Fig.\ref{fig:Domain}, it contributes to the norm only through the real positive prefactor $e^{\mathrm{Im}\eta}$.
Therefore, in the stable region defined by $\cos(2v)>0$, 
the norm is real and positive, while it diverges at the boundaries where $\cos(2v)=0$.

It follows from \eqref{varphiNorm} that the frame vacuum
$|\varphi_{\xi,0}\>$ is normalizable with a positive-definite
norm when the imaginary part $v$ lies in the stable domain
\begin{align}
-\frac{\pi}{4}+n\pi < v < \frac{\pi}{4}+n\pi \;, \qquad n\in\mathbb{Z},
\end{align}
whereas in the unstable region it cannot be interpreted as an ordinary Hilbert-space vector.
Instead, it should be regarded as a generalized vector in a rigged Hilbert space, as clarified below.
The boundaries at $v=\pm\pi/4+n\pi$ define the Stokes lines
\cite{Bender1993,BenderBook}, where the norm diverges.
As $v$ approaches a Stokes line from within the stable domain,
the normalization factor diverges as
$\propto(\pi/4+n\pi\mp v)^{-1/2}$,
indicating a characteristic square-root singularity 
associated with Stokes phenomena in non-compact quantum systems.

For the region with $\cos(2v)<0$, while $|\varphi_{\xi,0}\>$ is not normalizable in the Hilbert space, the frame vacuum is still well-defined as a generalized vector in the rigged-Hilbert space (RHS)~\cite{Gelfand1964,Petrosky1991a,Bohm1999,Chruscinski2003,Chruscinski2004,DeLaMadrid2005}. 
Indeed, corresponding to the right-vacuum $|\varphi_{\xi,0}\>$, we define the left-vacuum $\<\tilde\varphi_{\xi,0}|$ by 
\begin{align}
\<\tilde\varphi_{\xi,0}|\hat\varphi_\xi^\star=0 \;.
\end{align}
Corresponding to (\ref{vacuumdef}), we obtain
\begin{align}\label{vacuumdef2}
\<\tilde\varphi_{\xi,0}|=e^{i\eta/ 2}\<0|\hat V_2(\beta)\hat V_3(\theta) \;,
\end{align}
which likewise fails to belong to the Hilbert space in the unstable domain.
Thus, both right- and left-vacua belong to the RHS in the unstable domain\cite{DeLaMadrid2005}:
\begin{align}
|\varphi_{\xi,0}\> \in \Phi^\times \;,\; \<\tilde\varphi_{\xi,0}| \in \Phi' \;,
\end{align}
where $\Phi'$ and $\Phi^\times$ denote the dual and antidual spaces of the dense subspace $\Phi \subset \mathcal H$, respectively:
\begin{align}
\Phi \subset \mathcal H \subset \Phi^\times,
\qquad
\Phi \subset \mathcal H \subset \Phi'.
\end{align}
Here the bra is not understood as the Hermitian conjugate of the ket, but as an independent generalized eigenvector in the sense of the rigged Hilbert space formalism.
We may then define the normalization in the RHS  through the bi-orthonormal pairing \cite{Chruscinski2003,Chruscinski2004}:
\begin{align}
\< \tilde\varphi_{\xi,0} | \varphi_{\xi,0} \>
= \< 0 | 0 \> = 1 .
\end{align}

Figure~\ref{fig:Domain} illustrates this instability structure in the complex
$e^{\beta}$-plane.
The blank regions correspond to the stable domains where the frame vacuum
is normalizable in the Hilbert space, while the shaded regions represent
the unstable domains where the norm ceases to be real and the vacuum must be interpreted within the rigged-Hilbert-space framework.
The Stokes lines form the boundaries between these domains, along which the norm diverges, and act as
branch cuts separating different analytic sectors.
The exceptional points (black dots), to be characterized in the next section, occur in the limit $|u|\to\infty$ at discrete phase directions $v=n\pi/4$ $(n\in\mathbb Z)$.
In addition to the static phase diagram, we also indicate a representative
trajectory that passes from a stable region into an unstable one and
returns to another stable region.
The physical realization and interpretation of such trajectories will be
discussed in the next section.
We note that the complex-structure operator $J_\xi$ introduced in \eqref{appeq:Jmap}  carries a natural $\mathbb{Z}_4$ symmetry, since $J_\xi^4=\mathrm{id}$.
The origin of this fourfold structure will be clarified in the next section, where it will be shown that the squeezing parameter $\beta$ becomes a four-valued function of the generator parameters $(k_1,k_2,k_3)$.

These frame vacua in the RHS are the lowest bi-orthonormal eigenstates of a generator at the local frame $(\hat\varphi_\xi,\hat\varphi_\xi^\star)$ given by
\begin{align}\label{Kxi3}
&\hat K_{\xi,3}={1\over 2}\left(\hat\varphi_\xi^\star \hat\varphi_\xi+\hat\varphi_\xi \hat\varphi_\xi^\star \right) =\hat\varphi_\xi^\star \hat\varphi_\xi+\frac 12 \notag\\
&={1\over 2}\cosh2\beta\left (\hat a^\dagger \hat a+\hat a\hat a^\dagger \right)
+{1\over 2} \sinh2\beta\left(e^{-2i\theta}\hat a^{\dagger 2}+e^{2i\theta}\hat a^2\right) \;.
\end{align}
Since $\hat K_{\xi,3}$ is invariant under the gauge action generated by $\hat V_3(\eta)$, it defines the distinguished axis of the local frame.
We therefore call $\hat K_{\xi,3}$ the principal-axis frame generator.

The principal-axis frame generator determines the bi-orthonormal eigenbasis in the RHS associated with a local frame $(\hat \varphi_\xi,\hat\varphi_\xi^\star)$.
The complex eigenvalue problem of $\hat K_{\xi,3}$ in the RHS reads
\begin{align}
\hat K_{\xi,3}|\varphi_{\xi,n}\rangle
= z_n |\varphi_{\xi,n}\rangle \; ,\qquad
\langle\tilde\varphi_{\xi,n}|\hat K_{\xi,3}
= z_n \langle\tilde\varphi_{\xi,n}| \; ,
\end{align}
where the eigenvalues are given by $z_n=n+1/2$ with $n=0,1,2,\ldots$.
The eigenvectors satisfy
\begin{align}
\langle\tilde\varphi_{\xi,m}|\varphi_{\xi,n}\rangle
=\delta_{m,n}\;,\qquad
\sum_{n=0}^\infty
|\varphi_{\xi,n}\rangle
\langle\tilde\varphi_{\xi,n}|=1 \;.
\end{align}
The explicit forms of the eigenstates are given by (\ref{operatorRep})
\begin{subequations}\label{RHSstates}
\begin{align}
&|\varphi_{\xi,n}\>=e^{-i (n+1/2)\eta}\hat V_3^{-1}(\theta)\hat V^{-1}_2(\beta) |n\> \;,\\
&\<\tilde\varphi_{\xi,n}|=e^{i (n+1/2) \eta}\<n| \hat V_2(\beta)\hat V_3(\theta)  \;,
\end{align}
\end{subequations}
or equivalently,
\begin{align}\label{RHSstates2}
|\varphi_{\xi,n}\>={1\over \sqrt{n!}}\hat\varphi^{\star n}|\varphi_{\xi,0}\> \;,\;
\<\tilde\varphi_{\xi,n}|={1\over \sqrt{n!}}\<\tilde\varphi_{\xi,0}|\hat\varphi^{n}\;.
\end{align}

We have obtained the quantum geometric tensor for the RHS eigenstates with use of the conventional definition of the Berry connection for the non-Hermitian Hamiltonian~\cite{Garrison1988}, which we call the {\it state-space QGT}, while the QGT obtained in (\ref{QGTB}) in the preceding section is called operator-space QGT.
We compare the two QGTs  in Appendix~\ref{Appsec:QGTRHS}, where the state-space QGT is shown to be consistent with the operator-space QGT with the additional factors of the excitation number $n$.

We have shown that the frame vacuum exhibits a phase transition at the Stokes lines in terms of the divergence of the norm.
In contrast, the operator-space QGT derived in Sec.~\ref{sec:FrameTrans} is an analytic function of $\beta$, hence it can be continued across Stokes lines, which provides a new perspective for the frame vacuum phase transition.
This situation is closely analogous to an alternative interpretation
of quantum transitions based on analytic continuation into the complex-energy plane,
as studied in non-Hermitian and PT-symmetric quantum mechanics~\cite{Bender1993}.

Although the quantum geometric tensor generally takes complex values when
the squeezing parameter $\beta$ is treated as a complex variable,
both the quantum metric (symmetric part) and the Berry curvature (antisymmetric part),
given by \eqref{gmunu} and \eqref{Fmunu}, remain analytic functions of $\beta$.
As a consequence, the QGT varies smoothly across the frame-vacuum phase boundaries
(Stokes lines) without exhibiting any singular behavior.

This analyticity indicates that the frame-vacuum phase transition is not characterized
by a breakdown of the geometric structure itself, but rather by a change in the
realizability of the frame vacuum within the Hilbert space.
In the unstable domain, the vacuum ceases to be normalizable and must be described
within the framework of the rigged Hilbert space, while the underlying operator-space
geometry encoded in the QGT remains well defined.
In this sense, the analytically continued QGT provides a continuous geometric probe
that connects the stable and unstable domains across the Stokes lines.


\section{Physical Realization of the Frame Vacuum Phase Transition}\label{sec:Physical}

In the preceding sections, we have shown that the QGT defined on the operator-frame bundle is  analytic on the complex squeezing-parameter plane and varies smoothly  across the frame vacuum phase boundaries (Stokes lines).
We then ask what kind of realistic physical setting allows one to observe this continuous geometric change of the QGT on the complex parameter plane.
It should be noted that each point on the base manifold represents a distinct operator frame, which in turn defines its corresponding frame vacuum.
Therefore, we look for a physical generator that induces transitions between different operator frames, that is, transitions between the different frame vacua.
This situation is distinct from the Hamiltonian flow on a fixed classical phase space discussed in Sec.\ref{sec:CM}, where a specific Hamiltonian generates trajectories within a single, fixed phase space.
In the present case, the trajectory is defined on the operator-frame base manifold itself and represents a continuous transformation between different frames.
In this section, we illustrate a physical generator that induces such transitions between the different operator frames.

In order to construct a transition path on the base manifold through a physical operation, we consider the following Hermitian generator as a frame generator, which reduces to the principal-axis frame generator $\hat K_{\xi,3}$ introduced in the preceding section, upon  identifying the parameters as  shown in Eq.~(\ref{ksparameter}).
The purpose here is to consider a physical operation which drives the path into the complex parameter space, thereby providing a  connection between the stable and unstable domains.
To this end, in addition to the reference  mode  $(\hat a,\hat a^\dagger)$,  we introduce a coupling to an infinite number of environmental modes  $\hat b_k (\hat b_k^\dagger)$ with a continuous spectrum $\omega_k$ :
\begin{subequations}\label{Ktotal}
\begin{align}
\hat K({\bs k})&=k_3\hat K_3+k_1 \hat K_1+ k_2 \hat K_2 \\ 
&+ \int \omega_k \hat b_k^\dagger \hat b_k dk + \int g_k (\hat a^\dagger \hat b_k+\hat b_k^\dagger \hat a) dk \;,
\end{align}
\end{subequations}
where  ${\bs k}=(k_1,k_2,k_3)$ and  the last term represents the coupling between the reference-frame mode and the environment modes.
Note that $\hat K({\boldsymbol k})$ is Hermitian and represents a physical operation.
By comparison with Eq.~(\ref{Kxi3}), we see that the first line of $\hat K({\boldsymbol k})$
in Eq.~(\ref{Ktotal}) coincides with $\hat K_{\xi,3}$ for real $\beta$, provided that
\begin{align}\label{ksparameter}
k_3=\cosh2\beta \;, \; k_1=\sinh 2\beta \cos 2\theta\;,\; k_2=\sinh 2\beta \sin 2\theta\;.
\end{align}
Here we demonstrate how a real squeezing parameter $\beta$ can be effectively promoted to a complex value.
We note that this generator has been widely employed as a model Hamiltonian for quantum vacuum instability
in the presence of dissipation, including the dissipative dynamical Casimir effect,
dissipative parametric amplification, and dissipative Dicke-type models\cite{Milburn1981,Gardiner1984,Dimer2007,Ancheyata2017,Tanaka2020a,Dodonov2020}.

Once a frame generator is specified, it induces a continuous family of operator frames
along a path on the operator-frame base manifold.
To parametrize such a path, we introduce a real parameter $\lambda$,
which measures distance along the curve and should not be interpreted as physical time.

Let $\hat K(\mathbf{k}(\lambda))$ denote the frame generator defined in Eq.~\eqref{Ktotal},
with parameters $\mathbf{k}(\lambda)$ varying smoothly along the path.
The corresponding deformation of any operator $\hat{\mathcal O}$,
where $\hat{\mathcal O}=\hat a,\hat a^\dagger,\hat b_k,\hat b_k^\dagger$,
is given by
\begin{align}
-i\frac{d}{d\lambda}\hat{\mathcal O}
=
\mathfrak L_{\hat K}\hat{\mathcal O},
\end{align}
where $\mathfrak L_{\hat K}$ is regarded as a symplectic, Hamiltonian-type vector field
acting as a superoperator on the operator algebra.
With the sign convention adopted in this work, it is defined by
\begin{align}
\mathfrak L_{\hat K}\cdot := [\hat K,\cdot].
\end{align}
This convention is chosen so that, for the unperturbed generator
$\hat K=k_3\hat K_3$, one obtains
$\mathfrak L_{\hat K}\hat a=-k_3\hat a$ and
$\mathfrak L_{\hat K}\hat a^\dagger=k_3\hat a^\dagger$.
Thus the annihilation operator evolves as
$\hat a(\lambda)=e^{-ik_3\lambda}\hat a(0)$.
We shall call $\mathfrak L_{\hat K}$ the Liouvillian associated with $\hat K$,
in the sense that it acts as a superoperator on operators.
Here, the parameter $\lambda$ merely labels points along the path on the base manifold,
and the above equations describe a continuous deformation of the operator frame,
rather than the real-time evolution of a quantum state.
Accordingly, the resulting trajectory should be understood as a path on the
operator-frame base manifold generated by successive frame transformations,
and not as a Hamiltonian flow on a fixed phase space.

The matrix representation of the Liouvillian ${\mathfrak L}_{\hat K}$
is given by an infinite-dimensional matrix, as shown in Eq.~(\ref{Appeq:Lmat}).
By employing the Feshbach projection method~\cite{Feshbach1958,Petrosky1991a,Petrosky2000,Ordonez2001,Tanaka2020a},
we derive an effective Liouvillian in which the influence of the environmental modes
$(\hat b_k,\hat b_k^\dagger)$ on the reference-frame mode
$(\hat a,\hat a^\dagger)$ is incorporated through an environment-induced kernel
(commonly referred to as a self-energy).
The resulting effective Liouvillian, expressed in the reduced operator basis
$(\hat a,\hat a^\dagger)$, takes the form
\begin{align}\label{Leff}
{\bm L}_{\rm eff}(z;{\bs k})=
  \begin{pmatrix}
  -k_3 + \sigma_+(z;{\bs k}) & -k_1 + i k_2 \\
   k_1 + i k_2 &  k_3 + \sigma_-(z;{\bs k})
  \end{pmatrix} ,
\end{align}
where $\sigma_\pm(z;{\bs k})$ denote the environment-induced self-energy functions,
whose explicit forms are given in Appendix~\ref{Appsec:Leff}.
This reduction is exact in the sense that it preserves the Hermitian structure
of the full frame generator prior to projection, provided that the full
parameter dependence of the self-energy kernel is retained
\cite{Petrosky1991a,Petrosky2000,Ordonez2001,Tanaka2020a}.

Here, instead of retaining the dependence of $ \sigma(z;{\bs k}) $ on $ z $,
we introduce phenomenological parameters
$ \kappa $ and $ \gamma $ such that
$ \sigma(z;{\bs k}) \simeq \kappa + i \gamma $,
which is a standard approximation in the weak-coupling
regime of the system--environment interaction.
Within this approximation, the phenomenological effective Liouvillian takes the form
\begin{align}\label{Lphenom}
\bm L_{\rm ph}({\bs k})=
  \begin{pmatrix}
  -\mathcal K_3 & -k_1 + i k_2 \\
   k_1 + i k_2 &  \mathcal K_3
  \end{pmatrix} ,
\end{align}
where $ \mathcal K_3 := k_3 + \kappa + i \gamma $.
This effective Liouvillian represents the reduced symplectic (Hamiltonian-type)
vector field generated by an effective, generally non-Hermitian frame generator,
\begin{align}\label{Hph}
\hat K_{\rm ph}({\bs k})
= \mathcal K_3 \hat K_3 + k_1 \hat K_1 + k_2 \hat K_2 ,
\end{align}
in which the complexification of $ \mathcal K_3 $ induces a non-unitary
frame transformation of the mode operators.

The effective frame generator $\hat K_{\rm ph}({\bs k})$ is brought into a diagonal form
by the complex symplectic (Bogoliubov) transformation~(\ref{Sactonphi0}),
with the parametrization
\begin{align}\label{betaParam}
\beta=\frac 14 \ln\left({\C K}_3+k_\parallel\over {{\C K}_3-k_\parallel }\right) \;, \;\tan 2\theta ={k_2\over k_1}\;,\; k_\parallel:=\sqrt{k_1^2+k_2^2}\;,
\end{align}
which yields
\begin{align}
\hat K_{\rm ph}({\bs k})={Z_{\bs k} \over  2} \left(\hat\varphi_{\bs k}^\star\hat\varphi_{\bs k}+\hat\varphi_{\bs k}\hat\varphi_{\bs k}^\star\right) \;.
\end{align}
Here the eigenspectrum is given by
\begin{align}\label{Zval}
Z_{\bs k}=\sqrt{{\C K}^2_3-k_\parallel^2} \;.
\end{align}
It follows from Eq.~(\ref{Zval}) that $\mathcal K_3 = \pm k_\parallel$
defines exceptional points, at which the two operator modes
$(\hat\varphi_{\bs k}, \hat\varphi_{\bs k}^\star)$ coalesce and the spectral gap $Z_{\bs k}$
vanishes.
These exceptional points are defined at the level of the effective Liouvillian
and correspond to the coalescence of the canonical operator frame modes,
as discussed in Ref.~\cite{Tanaka2020a}.

\begin{figure}[t]
\begin{center}
\includegraphics[width=7cm]{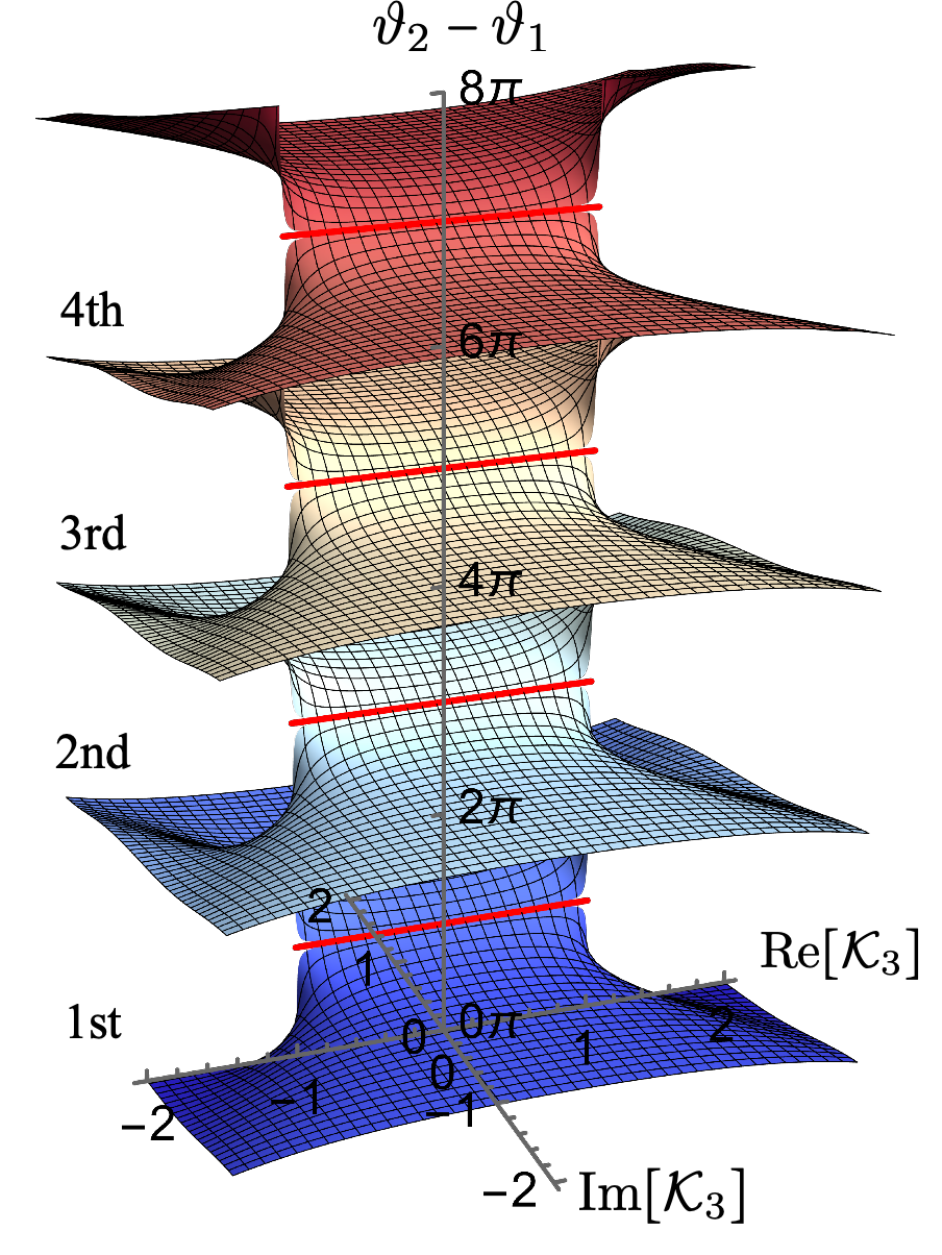}
\caption{
(Color online)
Argument of the complex squeezing parameter $e^{\beta}$, 
$4\,{\rm Im}\beta=\vartheta_2-\vartheta_1$, 
plotted on the complex ${\cal K}_3$ plane for $k_\parallel=1$.
The exceptional points at ${\cal K}_3=\pm k_\parallel$ appear as branch points,
around which the phase of $e^{\beta}$ exhibits a four-cycle ($\mathbb{Z}_4$) monodromy.
Each sector corresponds to a distinct Riemann sheet in the analytically continued parameter space.
The Stokes lines shown by the red lines, located at $\vartheta_2-\vartheta_1=\pi (2l+1)$ ($l\in\mathbb{Z}$),
emanate from the exceptional points and become branch cuts separating stable and unstable domains.
}\label{fig:angle}
\end{center}
\end{figure}

For comparison with the analysis in the preceding section, we define
\begin{align}\label{defK3}
{\C K}_3-k_\parallel=\rho_1 e^{i\vartheta_1}\;,\; {\C K}_3+k_\parallel=\rho_2 e^{i\vartheta_2}\;,
\end{align}
which leads to
\begin{align}\label{Compbeta}
\beta=\frac 14 \ln \left({\rho_2\over \rho_1}\right)+\frac i4 (\vartheta_2-\vartheta_1 ) \;.
\end{align}
The imaginary part of the complex squeezing parameter $v=(\vartheta_2-\vartheta_1)/4$ is controlled by the dissipation interaction of ${\rm Im}{\C K}_3$ as
\begin{align}
\vartheta_2-\vartheta_1=-\arctan\left( {2 k_\parallel {\rm Im} {\C K}_3 \over ({\rm Re}{\C K}_3)^2-(k_\parallel^2+({\rm Im}{\C K}_3)^2)} \right) \;.
\end{align}
Figure \ref{fig:angle} shows the argument of  $e^\beta$,  namely $4v=\vartheta_2-\vartheta_1$, in the complex ${\C K}_3$ plane for $k_\parallel=1$.

It follows  from (\ref{defK3}) that $\beta$ diverges at ${\C K}_3=\pm k_\parallel$, where $\rho_{1,2}=0$.
These points   correspond to the exceptional points shown by the black dots in Fig.\ref{fig:Domain}.
The ${\bb Z}_4$-monodromy associated with the complex structure illustrated in Fig.\ref{fig:Domain} is manifested  as the four-cycle periodic structure in Fig.\ref{fig:angle}, where the first through fourth  Riemann sheets  are indicated in correspondence to those in Fig.\ref{fig:Domain}.
The Stokes lines shown as red lines and appearing at
$\vartheta_2-\vartheta_1=\pi(2l+1)$ ($l\in\mathbb Z$)
divide each Riemann sheet and form branch cuts connecting
the exceptional points ${\cal K}_3=\pm 1$, which act as branch points of the analytically
continued parameter space.

\begin{figure}[t]
\begin{center}
\includegraphics[width=7cm]{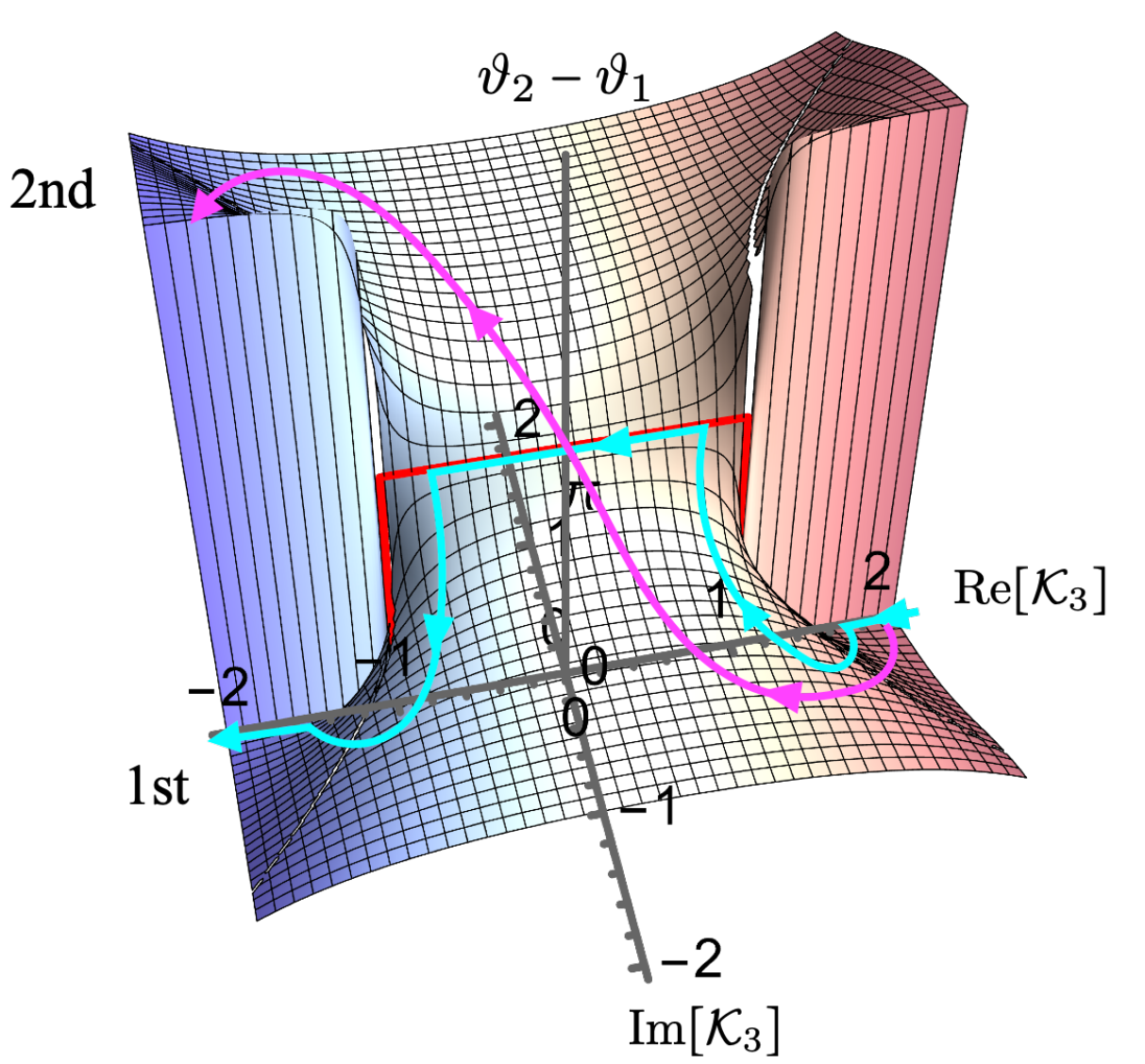}
\caption{
(Color online)
Continuous trajectories in the complex ${\cal K}_3$ plane encircling the exceptional points
at ${\cal K}_3=\pm k_\parallel$ on the first Riemann sheet: the blue curve runs on the first Riemann sheet, and the magenta curve runs from the first to the second Riemann sheets crossing over the Stokes line. 
A finite imaginary part ${\rm Im}{\cal K}_3$, induced by dissipation,
allows the system to bypass the exceptional points and cross the Stokes lines smoothly.
The corresponding trajectories are also drawn in  Fig.~\ref{fig:Domain}.
}\label{fig:Pathway}
\end{center}
\end{figure}

Without dissipation, i.e., ${\rm Im}\,\mathcal K_3 = 0$, 
the trajectory is confined to the line
$v = \arg(e^\beta) = \pi l/4$.
As a result, any transition from the stable to the unstable domain
necessarily passes through an exceptional point (EP),
as shown in Fig.~\ref{fig:Domain},
where the QGT diverges.
By contrast, when dissipation is introduced,
represented by a finite ${\rm Im}\,\mathcal K_3$,
the trajectory can circumvent the exceptional points
and continuously connect the stable and unstable regions.
Figure~\ref{fig:Pathway} illustrates two representative smooth trajectories:
one connecting the stable and unstable regions entirely within the first
Riemann sheet (blue curve),
and another crossing the Stokes line from the first to the second
Riemann sheet (magenta curve).
The corresponding paths are also indicated in Fig.~\ref{fig:Domain}.

Along these trajectories, the QGT obtained in the present work exhibits a continuous geometric evolution on the complexified operator-frame manifold.
This should be contrasted with the case in which the frame transformation is restricted to the real unitary domain.
In that case, the squeezing parameter $\beta$ diverges at the exceptional point, and the operator-frame transformation itself becomes singular.
After analytic continuation into the complex parameter space, however, the path can avoid the exceptional point and connect the stable and unstable domains through a smooth non-unitary operator-frame transformation.

The frame-vacuum transition then appears not as a singularity of the analytically continued operator-frame geometry, but as a change in the Hilbert-space realizability of the associated frame vacuum.
When the trajectory crosses a Stokes line, the frame vacuum ceases to be a normalizable Hilbert-space vector and must instead be interpreted within the rigged-Hilbert-space framework.
In this sense, the environmental coupling provides a concrete physical mechanism for realizing a smooth operator-frame deformation across a boundary where the corresponding Hilbert-space vacuum undergoes a frame-vacuum phase transition.

\section{Conclusion}\label{sec:Conclusion}

In this work, we have developed a geometric formulation of a non-compact quantum system,
focusing on a single-mode bosonic system in terms of a canonical operator frame bundle structure.
The canonical local operator frame naturally possesses a flat K\"ahler structure.
Complex symplectic transformations induce a nontrivial operator bundle structure with the
non-compact complex structure group ${\mathbb C}^\times$, reflecting the underlying gauge freedom.
The Ehresmann connection determines the parallel transport of the operator frame on the total space
of the frame bundle, and the parallel-transported canonical operators are obtained through a
covariant Taylor expansion.
We determine the Berry connection by imposing an appropriate gauge-fixing condition,
which specifies the parallel transport of the canonical operators on the total space
of the frame bundle.
The quantum geometric tensor (QGT) on the base manifold is then defined through
the line element associated with the infinitesimal parallel transport of the operator frame.

We have also constructed the frame vacuum associated with each local frame by transforming the bare
vacuum.
When the frame lies in the unstable domain, this transformation becomes non-unitary, and the frame
vacuum necessarily belongs to the rigged Hilbert space.
We illustrated the phase diagram characterizing the frame vacuum instability on the
complex-squeezing parameter plane, which is divided into four Riemann sheets by the Stokes lines.
The norm of the frame vacuum diverges at these Stokes lines.

We further showed that dissipative processes provide a realistic physical setting in which this
continuous geometric change at the frame vacuum phase transition can be observed.
By tuning the dissipation parameters, one can trace continuous pathways connecting the stable and unstable regions on the complex base manifold.
The resulting four-cycle Riemann surface structure originates from the complex structure of the operator frame, and we demonstrated that continuous trajectories encircling the exceptional points can be realized through controlled dissipation.

We would like to emphasize that in this work we conceptually separate the geometry of canonical operator frames from Hamiltonian dynamics. 
The frame vacuum and its instability are defined independently of any specific Hamiltonian.
We found that the frame vacuum instability does not correspond to a geometric singularity.
Rather, it reflects a breakdown of the Hilbert-space representation,
while the underlying geometric structure remains intact.
Therefore, the frame vacuum phase transition is regarded as a representational instability.
By this, we mean a breakdown of the Hilbert-space realization of the operator algebra and its vacuum, while the underlying geometric and algebraic structures remain well-defined and analytic.

Finally, we emphasize that the geometric change of the operator frame is not a consequence of dynamical evolution.
Rather, it arises from changes in the operator frame itself and the associated frame vacuum,
in a manner analogous in spirit to observer-dependent vacuum structures such as the Unruh effect\cite{Unruh1976,Nation2012,Capolupo2013}.
The relation between the present framework and the geometry of quantum systems on curved manifolds is an important subject for future investigation.

\acknowledgments
We thank K.~Nakao, H.~Ishihara, Y.~Kayanuma, M.~Morikawa, and J.~Inouye for valuable discussions.
This work was supported by JSPS KAKENHI Grant No.~24K06901.

\appendix

\section{A complex structure}\label{AppSec:Complex}

In \eqref{JonV} and \eqref{JonFr0}, we have shown the action of the complex structures $J_0$ and $\bs J_0$ on $V={\rm span}_{\bb C}\{\hat a,\hat a^\dagger\}$ and the reference frame $\hat{\bs\phi}_0$, respectively.
As stated in \S \ref{sec:Quant}, we show how the complex structures depend on the local operator frames.

We consider the complex structure $J_\xi$ on the local operator frame $\hat{\bs \phi}_\xi=(\hat\varphi_\xi,\hat\varphi_\xi^\star)$.
The operator vector space $V$ is equivalently spanned by the local operator frame
$\hat{\bs\phi}_\xi=(\hat\varphi_\xi,\hat\varphi_\xi^\star)$:
\begin{align}
V=\mathrm{span}_{\mathbb C}\{\hat\varphi_\xi,\hat\varphi^\star_\xi\} \;.
\end{align}
The action of  $J_\xi$ on $\hat{\bs\phi}_\xi$ is defined in the same way as in  \eqref{JonV} by
\begin{align}
J_\xi: V\to V \;,
\end{align}
which is represented as 
\begin{align}\label{appeq:Jmap}
J_\xi(\hat\varphi_\xi):=\hat\varphi^\star_\xi \;,\;
J_\xi(\hat\varphi^\star_\xi):=-\hat\varphi_\xi\;.
\end{align}
The action of the $J_\xi$ on the operator-frame $\hat{\bs\phi}_\xi$ is also defined by 
\begin{align}\label{Appeq:JonFrxi}
&\bs J_\xi : {\rm Fr}_\Omega(V) \to {\rm Fr}_\Omega(V)  \;,\notag\\
&{\bs J}_\xi(\hat{\bs \phi}_\xi) :=
\begin{pmatrix}
J_\xi(\hat\varphi_\xi) \\
J_\xi(\hat \varphi_\xi^\star)
\end{pmatrix}
=\begin{pmatrix}
\hat \varphi_\xi^\star \\
-\hat \varphi_\xi
\end{pmatrix} 
=
\begin{pmatrix}
0 & 1\\
-1 & 0
\end{pmatrix}
\begin{pmatrix}
\hat \varphi_\xi \\
\hat \varphi_\xi^\star
\end{pmatrix} \;.
\end{align}

Writing the linear symplectic transformation on $V$ as 
\begin{align}
S_\xi(\hat a)=\hat\varphi_\xi \;,\; S_\xi(\hat a^\dagger)=\hat\varphi_\xi^\star \;,
\end{align}
the transformation of the complex structures is given by
\begin{align}\label{Jtr}
J_\xi=S_{\xi}J_0 S_\xi^{-1} \;.
\end{align}
In fact, we find
\begin{subequations}
\begin{align}
&J_\xi(\hat\varphi_\xi)=S_{\xi}J_0 S_\xi^{-1} (\hat\varphi_\xi)=S_\xi J_0(\hat a)=S_\xi(\hat a^\dagger)=\hat \varphi_\xi^\star \;,\\
&J_\xi(\hat\varphi_\xi^\star)=S_{\xi}J_0 S_\xi^{-1} (\hat\varphi^\star_\xi)=S_\xi J_0(\hat a^\dagger)=S_\xi(-\hat a)=-\hat \varphi_\xi   \;,
\end{align}
\end{subequations}
which is consistent with \eqref{appeq:Jmap}.


\section{Ehresmann connection}\label{AppSec:Ehresmann}

In the present work, the principal bundle $P$ is identified with
$SL(2,\mathbb C)$ with structure group $\mathbb C^\times$, so that the base manifold is the homogeneous space $SL(2,\mathbb C)/\mathbb C^\times$.
This connection may equivalently be characterized by the reductive
decomposition of $\mathfrak{sl}(2,\mathbb C)$ induced by the nondegenerate
Killing form, which identifies the horizontal subspace with the complement
of $\mathrm{Lie}(\mathbb C^\times)$.
We briefly review the Ehresmann connection, which decomposes the tangent space at each point $u \in P$ into vertical and horizontal subspaces, characterized by the right-action covariance of the structure group $G$:
\begin{subequations}
\begin{align}
&T_u(P)=V_u (P)\oplus H_u (P) \;,\\
&R_{g*}H_u(P)=H_{ug}(P)\,\; \;,\; u \in P
\end{align}
\end{subequations}
where $R_g$ denotes the right action of  $g\in G$.

In the Ehresmann connection, the horizontal space is defined by a connection one-form $\omega \in \Omega^1(P,{\mathfrak g})$ satisfying
\begin{subequations}
\begin{align}
&\omega(A^\sharp) = A \in {\mathfrak g}, \quad \text{for} \;A^\sharp \in V_uP  \\
&R_g^*\omega = \mathrm{Ad}_{g^{-1}} \,\omega   \;,
\end{align}
\end{subequations}
where $A^\sharp$ is the fundamental vertical vector field associated with $A\in {\mathfrak g}$.
The horizontal tangent space is then defined as the kernel of $\omega$:
\begin{align}
\omega(X_u) = 0, \; \;  \text{for} \; X_u \in H_uP  \;.
\end{align}

Locally, the connection one-form can be expressed as
\begin{align}
\omega=\zeta_i(u) ^{-1} \pi ^*\omega_i  \zeta_i(u)  +  \zeta_i ^{-1}(u) d_P \zeta_i(u) \;,
\end{align}
where $\zeta_i(u)\in G$ is a local trivialization function on $P$, and $\omega_i$ is a ${\mathfrak g}$-valued one-form on an open set $U_i \subset B$:
\begin{align}\label{AppEq:omegai}
\omega_i= A_{i,\mu}({\bs x})dx^\mu \; \in {\mathfrak g}\otimes T^*B \;,\; A_{i,\mu}({\bs x}) \in {\mathfrak g} \;,
\end{align}
where $A_{i,\mu}({\bs x})$ is regarded as the Berry connection or gauge potential.

A tangent vector on $B$ is horizontally lifted to $P$ as
\begin{align}
\partial_\mu \longmapsto
\tilde\partial_\mu = \sigma_{i*}\partial_\mu - A^\sharp_{i,\mu},
\end{align}
which satisfies the horizontality condition  $\omega(\tilde\partial_\mu)=0 $.
Here we have adopted the standard local trivialization
$\zeta_i(u)=e$, so that $\omega=\pi^*\omega_i $,
and the vertical component  $A^\sharp_{i,\mu}$ precisely cancels
$\omega(\sigma_{i*}\partial_\mu) $.

This defines the horizontal lift of a curve, ensuring that its trajectory in the total space $P$ remains entirely within the horizontal subspace:
 \begin{align}
 {d\over dt} \gamma ({\bs x}(t)) \longmapsto {d\over dt} \tilde\gamma ({\bs x}(t),\tilde\zeta(t)) \; ;
 \end{align}
 This is precisely the definition of parallel transport on $P$, which uniquely determines the relation between the local trivialization $\tilde\zeta(t)$ and the Berry connection $A_\mu(t)$.
\section{Preservation of the Canonical Commutation relation}\label{AppSec:CCR}

In this section, we prove \eqref{CCRparallel}.

First, we introduce the induced covariant derivative
$\bs \nabla_\mu$ acting on composite operator expressions,
defined from the component covariant derivatives
$D_\mu$ and $D_\mu^\star$
through the Leibniz rule:
\begin{align}\label{DLeibniz}
\bs \nabla_\mu
(\hat\varphi_e \hat\varphi_e^\star)
:=
(D_\mu \hat\varphi_e)\hat\varphi_e^\star
+
\hat\varphi_e(D_\mu^\star \hat\varphi_e^\star).
\end{align}

In \eqref{CovariantD}, the opposite signs 
 ensure the preservation of the CCR structure under the action of $\bs \nabla_\mu$:
Using \eqref{DLeibniz}, we see 
\begin{align}\label{DmuAction}
{\bs \nabla}_\mu [\hat\varphi_e,\hat\varphi_e^\star]
=[D_\mu\hat\varphi_e, \hat\varphi_e^\star]+[\hat\varphi_e, D_\mu^\star\hat\varphi_e^\star]=
\partial_\mu [\hat\varphi_e,\hat\varphi_e^\star]
=0 \;.
\end{align}

We now prove  Eq.\eqref{CCRparallel}.
Suppose
\begin{equation}
C(x,t):=[\hat\varphi_D(x+dx,t),\hat\varphi^\star_D(x+dx,t)] \;,
\end{equation}
where we define
\begin{align}
&\hat\varphi_D(x+dx, t):=e^{t D_\mu dx^\mu}\hat\varphi(x) \;, \notag\\
&\hat\varphi^\star_D(x+dx,t):= e^{t D^\star_\mu dx^\mu}\hat\varphi^\star(x) \; .
\end{align}
We prove that
\begin{equation}\label{Cxt}
C(x,t)=C(x,0)=1  \;,\; \forall x \in B \;,\; \forall t\in {\bb R}\;.
\end{equation}

\begin{proof}
We first derive the partial differential equation of $C(x,t)$,
\begin{subequations}\label{dCt}
\begin{align}
{\partial\over \partial t}C(x,t)
&=\left[\frac \partial{\partial t}\hat\varphi_D(x+dx,t),\hat\varphi^\star_D(x+dx,t)\right] \notag\\
& \quad +\left[\hat\varphi_D(x+dx,t),\frac \partial{\partial t}\hat\varphi_D^\star(x+dx,t)\right] \\
&=dx^\mu\Big( \left[D_\mu  \hat\varphi_D(x+dx,t), \hat\varphi_D^\star(x+dx,t)\right]  \notag\\
&\quad+\left[\hat\varphi_D(x+dx,t), D_\mu^\star \hat\varphi_D^\star(x+dx,t)\right]  \Big)  \\
&=dx^\mu {\bs \nabla}_\mu[\hat\varphi_D(x+dx,t),\hat\varphi_D^\star(x+dx,t)]\\
&=dx^\mu {\bs \nabla}_\mu C(x,t)   \;.
\end{align} 
\end{subequations}
with the initial condition 
\begin{equation}\label{C0}
C(x,0)=1\;.
\end{equation}
In (\ref{dCt}d), we have used \eqref{DmuAction}.

Noticing that $\bs \nabla_\mu$ does not depend on $t$, $C(x,t)$ is formally solved by
 \begin{align}\label{solution}
 C(x,t) = e^{ t \bs \nabla_\mu dx^\mu}C(x,0) \;.
 \end{align}
If we assume
\begin{align}\label{CxtAssume}
C(x,t)=C(x,0) = 1=[\hat\varphi(x),\hat\varphi^\star(x)] \;,\; \forall t \in {\bb R} \;,
\end{align}
we find that this satisfies \eqref{dCt}:
\begin{align}
&\text{the l.h.s.} = {\partial \over \partial t} C(x,t)=0 \\
&\text{the r.h.s.}= dx^\mu \bs \nabla_\mu C(x,t)=dx^\mu \bs \nabla_\mu C(x,0) \notag \\
& \hspace{1cm}=dx^\mu[D_\mu\hat\varphi(x),\hat\varphi^\star(x)]+[\hat\varphi(x),D_\mu^\star\hat\varphi^\star(x)]
=0 \;,
\end{align}
where the last equality follows from Eq.\eqref{CovariantD}.

 Thus, $C(x,t)=1$ satisfies Eq.~\eqref{dCt} together with the initial condition \eqref{C0}. 
 By the uniqueness theorem for first-order differential equations, this solution is unique, proving Eq.~\eqref{Cxt}.

\end{proof}

\section{Intertwining relation}\label{AppSec:Intertwining}

In this section we derive \eqref{QGTB}.
For this purpose, we first prove the intertwining relation between $\bs J_e$ and $\bs D_\mu$:
\begin{align}\label{first}
{\bs J}_e\exp\left[ t  \bs D_\mu  dx^\mu\right] \hat{\bs \phi_e}(x)
=\exp\left[  t \bs D^\star_\mu  dx^\mu\right] \bs J_e  \hat{\bs \phi}_e(x) \;, \; \forall t \in {\bb R}\;,
\end{align}
where $\bs J_e$ acts on the  frame $\hat{\bs\phi}_e(x)$ as
\begin{align}
\bs J_e(\hat{\bs\phi}_e (x))
=\bs J_e 
\begin{pmatrix}
\hat\varphi_e(x) \\ \hat\varphi_e^\star(x)
\end{pmatrix}
=
\begin{pmatrix}
\hat\varphi_e^\star(x) \\ -\hat\varphi_e(x)
\end{pmatrix} \;,
\end{align}
corresponding to \eqref{JonFr0}.

\begin{proof}
The l.h.s. reads
\begin{align}
&{\bs J}_e\exp\left[ t  \bs D_\mu  dx^\mu\right] \hat{\bs \phi}_e(x) 
={\bs J}_e\exp\left[   \begin{pmatrix}D_\mu & 0 \\ 0  & D_\mu^\star \end{pmatrix} t dx^\mu\right]
\begin{pmatrix}\hat\varphi_e(x) \\\hat\varphi_e^\star(x) \end{pmatrix}  \notag\\ 
&={\bs J}_e\begin{pmatrix}\hat\varphi_D(x+t dx) \\  \hat\varphi_D^\star(x+t dx) \end{pmatrix} 
=\begin{pmatrix}\hat\varphi_D^\star(x+ t dx) \\ - \hat\varphi_D(x+t dx) \end{pmatrix}  \notag\\
&=\exp\left[ \begin{pmatrix}D^\star_\mu & 0 \\ 0  & D_\mu \end{pmatrix} t dx^\mu\right]
\begin{pmatrix}\hat\varphi_e^\star(x) \\ - \hat\varphi_e(x) \end{pmatrix} \notag \\
&=\exp\left[ \bs D^\star_\mu t dx^\mu\right]{\bs J}_e
\begin{pmatrix}\hat\varphi_e(x) \\  \hat\varphi_e^\star(x) \end{pmatrix} 
=\exp\left[ \bs D^\star_\mu  t dx^\mu\right] \bs J_e  \hat{\bs \phi}_e(x) \;,
\end{align}
where we have used \eqref{CCRparallel} to regard
$(\hat\varphi_D(x+tdx),\hat\varphi_D^\star(x+tdx))^T$
as an operator frame on which $\bs J_e$ acts.
This proves \eqref{first} 
\end{proof}

Next, differentiating Eq.~\eqref{first} with respect to $t$ and evaluating at $t=0$, we have
\begin{align}
\frac d {dt}\Big|_{t=0}{\bs J}_e\exp\left[ t  \bs D_\mu  dx^\mu\right] \hat{\bs \phi}(x)
=\frac d {dt}\Big|_{t=0}\exp\left[  t \bs D^\star_\mu  dx^\mu\right] \bs J_e  \hat{\bs \phi}(x) \;.
\end{align}
 This results in
\begin{align}
{\bs J}_e  \bs D_\mu  \hat{\bs \phi_e}(x) dx^\mu
= \bs D^\star_\mu   \bs J_e  \hat{\bs \phi}_e(x) dx^\mu
\end{align}
which proves
\begin{align}\label{Appeq:inter}
{\bs J}_e  \bs D_\mu  \hat{\bs \phi}_e(x) 
= \bs D^\star_\mu   \bs J_e  \hat{\bs \phi}_e(x)  \;.
\end{align}

Then the quantum geometric tensor given by \eqref{QGTdef} reads
\begin{subequations}\label{Appeq:GVdeviation}
\begin{align}
Q_{\mu\nu}&=-G_e(D_\nu \hat\varphi_e,D_\mu \hat\varphi_e) 
=\Omega(J_e D_\mu \hat\varphi_e, D_\nu \hat\varphi_e ) \notag\\
&=\Omega(D_\mu^\star J_e \hat\varphi_e, D_\nu \hat\varphi_e )\\
&=\Omega(D_\mu^\star \hat\varphi^\star_e , D_\nu \hat\varphi_e)
=[D_\mu^\star \hat\varphi^\star_e, D_\nu \hat\varphi_e ] \;.
\end{align}
\end{subequations}
where  in (\ref{Appeq:GVdeviation}a) we have used the intertwining relation \eqref{Appeq:inter}.


\section{QGT for the state vector in the rigged Hilbert space}\label{Appsec:QGTRHS}

In this section, we derive the QGT for the eigenstates of the principal-axis generator $\hat K_{\xi,3}$ in terms of the rigged Hilbert space.

For the left- and the right-eigenstates given by (\ref{RHSstates}), the QGT is defined by
\begin{align}
Q_{\xi;\mu\nu}:
=\<\partial_\mu\tilde\varphi_{\xi,n}|(1-|\varphi_{\xi,n}\>\<\tilde\varphi_{\xi,n}|)|\partial_\nu\varphi_{\xi,n}\> \;.
\end{align}
Using (\ref{RHSstates}), one obtains after direct calculation the  quantum metric $g_{\xi;\mu\nu}:=1/2 (Q_{\xi;\mu\nu}+Q_{\xi;\nu\mu})$ and the Berry curvature $F_{\xi;\mu\nu}:=(-i)(Q_{\xi;\mu\nu}-Q_{\xi;\nu\mu})$ as follows
\begin{subequations}\label{Appeq:gF}
\begin{align}
&g^{(n)}_{\xi;\mu\nu}=\frac 12
\begin{pmatrix}
 \sinh^2 2\beta &0 \\
 0 & 1
\end{pmatrix}
(n^2+n+1) \;,\\
\vspace{1pt}\notag\\
&F^{(n)}_{\xi;\mu\nu}=2
\begin{pmatrix}
0 &   \sinh 2\beta \\
 -   \sinh 2\beta &0
\end{pmatrix}
\left(n+\frac 12\right)  \;.
\end{align}
\end{subequations}

With use of the $\mathfrak {su}(1,1)$-generators for the local frame $(\hat\varphi_\xi,\hat\varphi_\xi^\star)$ defined by
\begin{align}
\hat K_{\xi,3} = \frac 12 \left(\hat \varphi_\xi \hat \varphi_\xi^{\star} 
+ \hat \varphi^{\star} \hat \varphi_\xi \right) ,\hat K_{\xi,+} =  \hat \varphi_\xi^{\star 2} ,
\hat K_{\xi,-} =\hat \varphi_\xi^{2} \;,
\end{align}
whose explicit transformation from the bare generators is given by
\begin{widetext}
\begin{align}
\label{frameGenerator}
\begin{pmatrix}
\hat K_{\xi,+}\\[2pt]
\hat K_{\xi,-}\\[2pt]
\hat K_{\xi,3}
\end{pmatrix}
=
\begin{pmatrix}
e^{-2i\eta} e^{-2i\theta}\cosh^2\beta
& e^{-2i\eta} e^{2i\theta}\sinh^2\beta
& e^{-2i\eta}\sinh 2\beta\\[2pt]
e^{2i\eta} e^{-2i\theta}\sinh^2\beta
& e^{2i\eta} e^{2i\theta}\cosh^2\beta
& e^{2i\eta}\sinh 2\beta\\[2pt]
\frac{1}{2} e^{-2i\theta}\sinh 2\beta
& \frac{1}{2} e^{2i\theta}\sinh 2\beta
& \cosh 2\beta
\end{pmatrix}
\begin{pmatrix}
\hat K_{+}\\[2pt]
\hat K_{-}\\[2pt]
\hat K_{3}
\end{pmatrix}\;,
\end{align}
\end{widetext}
the quantum metric and the Berry curvature  for the bi-orthogonal eigenstates are represented by
\begin{subequations}\label{Appeq:gBxi}
\begin{align}
&g^{(n)}_{\xi;\mu\nu}=\frac 12\;g_{\mu\nu} \<\tilde\varphi_{\xi,n}|\{\hat K_{\xi,-},\hat K_{\xi,+}\}_+|\varphi_{\xi,n}\>\;,\\
&F^{(n)}_{\xi;\mu\nu}=\frac12 F_{\mu\nu} \<\tilde\varphi_{\xi,n}| [\hat K_{\xi,-},\hat K_{\xi,+}] |\varphi_{\xi,n}\>\;,
\end{align}
\end{subequations}
where $\{\cdot,\cdot\}_+$ and $[\cdot,\cdot]$ are the  anticommutator (Jordan product)  and  the commutator (Lie bracket), respectively.
In \eqref{Appeq:gBxi}, $g_{\mu\nu}$ and $F_{\mu\nu}$ are the operator-space quantum metric and Berry curvature given by (\ref{gmunu}) and (\ref{Fmunu}), respectively.
The relations \eqref{Appeq:gBxi} show that the quantum metric and the Berry curvature
are governed by the symmetric and antisymmetric expectation values of the
two-particle excitation operators $\hat K_{\xi,-}$ and $\hat K_{\xi,+}$, respectively\cite{Ashtekar1999a}.
The quantum metric is associated with the fluctuation susceptibility
encoded in the symmetric (Jordan) product,
while the Berry curvature represents an antisymmetric geometric response
originating from the non-commutativity of these operators.

We note that the state-QGT is consistent with the operator-space QGT in its functional dependence on the squeezing parameter, while it additionally depends on the excitation number $n$.
This structure indicates that the operator-space QGT provides the universal
geometric sector, while the state-space QGT additionally contains
representation-dependent information through the expectation values of the
local frame generators.
Accordingly, the common $2\times2$ matrix structure shared by the operator-space
and state-space QGTs reflects an intrinsic geometric structure of the
operator-frame bundle itself, whereas the additional scalar factors encode
state-dependent dynamical information.

\section{Derivation of the effective Liouvillian in terms of the projection method}\label{Appsec:Leff}

In this section we derive the effective Liouvillian starting with the frame generator (\ref{Ktotal}) representing the coupling of the reference mode with the continuous environment modes.
Adding continuous environmental modes $\{\hat b_k\}$ and $\{\hat b_k^\dagger\}$
to the reference mode $(\hat a,\hat a^\dagger)$, the total operator basis is given by
\begin{align}
\left(\hat a, \{\hat b_k\}, \hat a^\dagger, \{\hat b_k^\dagger\}\right).
\end{align}

Hereafter, we use $\hat b$ and $\hat b^\dagger$ to represent the environmental
modes collectively, suppressing the mode index $k$ for notational simplicity.
Then, the quadratic  generator of the total system  is symbolically represented by
\begin{align}
\hat K=\frac 12
\left(\hat a, \hat b, \hat a^\dagger, \hat b^\dagger\right)
{\bs K} 
\left(\hat a, \hat b, \hat a^\dagger, \hat b^\dagger\right)^T
\end{align}
where ${\bs K}$ is the representation matrix of the quadratic generator
in the operator basis $(\hat a, \hat b, \hat a^\dagger, \hat b^\dagger)$,
\begin{equation}
{\bs K}=
\left(
\begin{array}{cccc}
 k_1+i k_2 & 0 &k_3 & \bar g\\
 0 & 0 & g & \omega \\
 k_3 & g & k_1- i k_2 &0 \\
 \bar g & \omega & 0 & 0
\end{array}
\right).
\end{equation}
Corresponding matrix representation of the Liouvillian is given by
\begin{align}\label{Appeq:Lmat}
{\bs L}
=-\bs\Omega {\bs K}
=
\begin{pmatrix}
-k_3 & -g  & -k_1 +i k_2 & 0\\
 -\bar g & -\omega  & 0 & 0\\
 k_1+i k_2 & 0 & k_3 & \bar g\\
 0 & 0  & g & \omega
\end{pmatrix}\;,
\end{align}
which is confirmed to satisfy the complex symplectic symmetry
\begin{align}
\bs\Omega {\bs L}\bs\Omega ={\bs L}^T \;,
\end{align}
which reflects the complex symplectic symmetry of the Liouvillian.
Changing the order of the operator basis to 
\begin{align}
 \left(\hat a, \hat a^\dagger, \hat b^\dagger,\hat b \right) 
\end{align}
brings about
\begin{align}
{\bs L}
=
\left(
\begin{array}{cc|cc}
-k_3 & -k_1 +i k_2  & -g  &  0\\
 k_1+i k_2 & k_3  & 0 & \bar g\\\cline{1-4}
-\bar g & 0 & -\omega & 0\\
 0 & g & 0 & \omega
\end{array}\right) \;,
\end{align}
where the upper-left block acts on the reduced operator subspace
$(\hat a,\hat a^\dagger)$.

To eliminate the environmental degrees of freedom and obtain an effective
generator acting only on the central mode, we employ the Feshbach-projection
method\cite{Petrosky1997,Chruscinski2013,Tanaka2020a}.
We consider the following projection operators whose representation matrices are given by
\begin{align}
{\bs P}
=\left(
\begin{array}{cc|cc}
1 & 0  & 0  &  0\\
 0 & 1  & 0 & 0 \\ \cline{1-4}
0 & 0 & 0 & 0\\
 0 & 0 & 0 & 0
\end{array}
\right) \;,\;
{\bs Q}
=\left(
\begin{array}{cc|cc}
0 & 0  & 0  &  0\\
 0 & 0  & 0 & 0 \\ \cline{1-4}
0 & 0 & 1 & 0\\
 0 & 0 & 0 & 1
\end{array}
\right) \;,\;
\end{align}
where ${\bs P}$ and ${\bs Q}$ represent the projector onto the central  and the environmental subspaces, respectively.

With use of the Feshbach-projection method, the eigenvalue-dependent effective Liouvillian is given by
\begin{align}\label{Appeq:Leff}
{\bs L}_{\rm eff}(z)&={\bs P} {\bs L} {\bs P}+ {\bs P} {\bs L} {\bs Q}{1\over z-{\bs Q}{\bs L}{\bs Q}}{\bs Q}{\bs L}{\bs P}\notag\\ 
&=
\begin{pmatrix}
-k_3 & -k_1+ik_2 \\
 k_1+ik_2 & k_3 
\end{pmatrix}
+
\begin{pmatrix}
{|g|^2 \over z+\omega} & 0\\
0 & {|g|^2 \over z-\omega}
\end{pmatrix}
\end{align}
where the terms $|g|^2/(z+\omega)$ and $|g|^2/(z-\omega)$ are symbolic expressions
of the $z$-dependent self-energy kernel defined by the Cauchy integral
\begin{align}\label{Appeq:self}
\sigma(z)_\pm={|g|^2\over z\pm \omega}:=\int {|g_k|^2\over z\pm \omega_k} dk \;.
\end{align}
Thus, (\ref{Appeq:Leff}) and (\ref{Appeq:self}) explain (\ref{Leff}).
Although the frame generator of the total system is Hermitian, the projected effective Liouvillian generally acquires a non-Hermitian structure through the analytically continued self-energy kernel, reflecting irreversible dissipation into the continuum\cite{Petrosky2000,Ordonez2001}.


\end{document}